\documentclass[a4paper,11pt]{article}
\usepackage{jheppub} 
\pdfoutput=1
\usepackage{graphicx,array}
\usepackage[dvipsnames]{xcolor}
\usepackage{amstext}
\usepackage{amssymb}
\usepackage{slashed}
\definecolor{orange}{rgb}{1,0.5,0}
\definecolor{brown}{rgb}{0.59, 0.29, 0.0}
\definecolor{note_fontcolor}{rgb}{0.80078125, 0.80078125, 0.80078125}

\newcommand{\LSF}[1]{\text{LSF}_\text{#1}}

\def\beq{\begin{equation}}
\def\eeq{\end{equation}}
\def\bea{\begin{eqnarray}}
\def\eea{\end{eqnarray}}

\usepackage{subcaption}
\title{Detecting a Boosted Diboson Resonance}

\author[a]{Kaustubh Agashe,}
\author[a,b]{Jack H.~Collins,}
\author[a]{Peizhi Du,}
\author[c]{Sungwoo Hong,}
\author[d,e]{Doojin Kim}
\author[f]{and Rashmish K.~Mishra}

\affiliation[a]{Maryland Center for Fundamental Physics, Department of Physics, University of Maryland, College Park, MD 20742, USA}
\affiliation[b]{Department of Physics and Astronomy, Johns Hopkins University, Baltimore, MD 21218, USA}
\affiliation[c]{Department of Physics, LEPP, Cornell University, Ithaca NY 14853, USA}
\affiliation[d]{Theoretical Physics Department, CERN, Geneva, Switzerland}
\affiliation[e]{Department of Physics, University of Arizona, Tucson, AZ 85721, USA}
\affiliation[f]{INFN, Pisa, Italy and Scuola Normale Superiore, Piazza dei Cavalieri 7, 56126 Pisa, Italy}
      
\emailAdd{kagashe@umd.edu}
\emailAdd{jhc296@umd.edu} 
\emailAdd{pdu@umd.edu} 
\emailAdd{sh768@cornell.edu} 
\emailAdd{doojinkim@email.arizona.edu} 
\emailAdd{rashmish@pi.infn.it}

\abstract{
New light scalar particles in the mass range of hundreds of GeV, decaying into a pair of $W/Z$ bosons can appear in several extensions of the SM. The focus of collider studies for such a scalar is often on its direct production, where the scalar is typically only mildly boosted. The observed $W/Z$ are therefore well-separated, allowing analyses for the scalar resonance in a standard fashion as a low-mass diboson resonance.
In this work we instead focus on the scenario where the direct production of the scalar is suppressed, and it is rather produced via the decay of a significantly heavier (a few TeV mass) new particle, in conjunction with SM particles. Such a process results in the scalar being highly boosted, rendering the $W/Z$'s from its decay merged. The final state in such a decay is a ``fat'' jet, which can be either four pronged (for fully hadronic $W/Z$ decays), or may be like a $W/Z$ jet, but with leptons buried inside (if one of the $W/Z$ decays leptonically). In addition, this fat jet has a jet mass that can be quite different from that of the $W/Z$/Higgs/top quark-induced jet, and may be missed by existing searches. 
In this work, we develop dedicated algorithms for tagging such multi-layered ``boosted  dibosons'' at the LHC.
As a concrete application, we discuss an extension of the standard warped extra dimensional framework where such a light scalar can arise. We demonstrate that the use of these algorithms gives sensitivity in mass ranges that are otherwise poorly constrained.
}

\begin{document} 

\begin{flushright}
UMD-PP-018-05 \\
CERN-TH-2018-194
\end{flushright}

\maketitle
\flushbottom

\section{Introduction}
\label{sec:introduction}

The discovery of the standard model (SM)-like Higgs boson with a mass of 125 GeV motivates searches for other scalar particles in the mass range of a few hundreds of GeV, with similar couplings. Such scalar particles can arise in several beyond the SM (BSM) scenarios, and have been studied extensively in the literature, both experimentally~\cite{CMS:2016jpd, CMS:2016noo} and theoretically~\cite{Branco:2011iw}. In most cases the emphasis has been on direct single production of such a scalar, often via gluon fusion, much like the SM Higgs itself. Such a process leads to the scalar being produced close to threshold with typically only a mild boost. Consequently, the decay products of the produced scalar are sufficiently isolated from each other and thus it is rather straightforward to identify the resulting signatures.

However, as we will illustrate below, it is possible to have other well motivated BSM scenarios where the couplings of the scalar to gluons and fermions can be small, while the couplings to electroweak (EW) bosons are still sizable. Due to the small couplings to the fermions and gluons, direct single production of the scalar is highly suppressed. Further, since the $W/Z$ content of the proton is small (relative to quarks/gluons), the direct production via vector boson fusion (VBF) can be small even if the scalar's interactions with $W/Z$ bosons are not small. Similarly, the rate for scalar production in association with $W/Z$ also can be small. In short, the rate for direct production of the scalar gets suppressed and consequently the bounds on it get weak. This is an interesting and viable variation relative to the case of the SM Higgs-like scalar.

In such a scenario, the leading production channel for such a light scalar particle (henceforth called $\varphi$) can be via {\em decays} of heavier BSM particles with masses in the TeV range, in association with other SM/BSM particles. The assumption is that the heavier parent particle might have unsuppressed couplings to quarks/gluons inside the proton, unlike $\varphi$. Hence the light scalar $\varphi$ is only accessible after paying the energy barrier price of the parent. A remarkable feature of this production mode for $\varphi$ is that it is highly boosted and thus its decay products are highly collimated. In the case that $\varphi$ decays into pairs of $W/Z$ bosons one might get a four-prong ``fat'' jet if both of these decay hadronically. Alternately, if only one of $W/Z$'s decays leptonically, one has a two-pronged jet with additional lepton(s) ``buried'' inside it. These signatures are possible for a boosted SM Higgs boson decaying into $WW^*$, as has been studied in the CMS search of Ref.~\cite{Khachatryan:2015bma} involving the four-pronged hadronic decay. However, in the context of a general BSM scenario the mass of the scalar is not fixed a priori, and it is a possibility that this scalar may be simply missed in existing searches.

\begin{figure}[t]
\centering
\includegraphics[width=9cm]{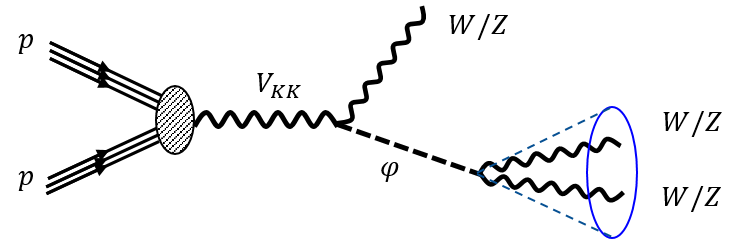}
\caption{
\small{The benchmark process under consideration, based on an extension of the warped extra-dimensional RS models. $V_{\rm KK}$ denotes the KK mode of $W/Z$, and corresponds to the heavier parents. $\varphi$ denotes the scalar, whose role is played by the radion. The cone around the two $W/Z$'s from the $\varphi$ decay delineates that they are highly collimated.}
}
\label{fig:scenario}
\end{figure}

It is therefore important to identify such scenarios and develop methods to isolate signals with such boosted topologies. In this paper, we study dedicated strategies for identifying such a light scalar decaying into merged $W/Z$ pairs, dubbed a ``boosted diboson'' resonance. As a specific application of this general strategy, we consider an extension of the standard warped extra dimensional setup~\cite{Agashe:2016rle}, where such a study is applicable. In this model, the role of $\varphi$ is played by the radion (the modulus corresponding to fluctuations in extra dimensional size), with its parent being the Kaluza-Klein (KK) excitation of a SM EW gauge boson, and accompanying SM particle being an EW gauge boson (see Fig.~\ref{fig:scenario}). We remark that in the standard warped setup consisting of a single region of bulk bounded by two branes, the rate for this entire process is rather small. On the contrary, what makes it large here is the presence of an extra region of bulk beyond the usual case (see~\cite{Agashe:2016rle} for more details).

We would like to emphasize here that our considerations are general -- the studied BSM scenario is mere illustrative, and the tagging method is suitable in general for multilayered boosted objects. For example, the strategies we employ should apply to other BSM models with a similar topology, even if the Lorentz structure of the couplings involved is different. In models with extended EW gauge sector (in four dimensions), $\varphi$ can be one of the scalars associated with breaking of the larger gauge symmetry and can be produced in decays of the associated $W^{ \prime }/Z^{ \prime }$ (see e.g. Ref.~\cite{Mohapatra:2016twe}). In such models, $\varphi$ couples to the vector field itself. This should be contrasted with the extra-dimensional model discussed above, where it couples to the field strength. We refer to these two kinds of coupling as ``vector'' and ``tensor'' respectively. Note that a coupling to photons must be of a tensor form because of gauge invariance.

Similar ideas have been discussed previously in the literature. For example, Ref.~\cite{Brust:2014gia} already made the general point that light new particles could be dominantly produced via decay of heavier ones, thus the light particles could be boosted, resulting in non-isolated SM particles from their decay. Specifically, the case of light scalars decaying into merged $W/Z$ pairs was studied in Refs.~\cite{Son:2012mb, Aguilar-Saavedra:2017zuc}. In addition to reiterating these earlier messages, our paper can be considered as a more systematic and complete investigation along these lines. For example, we analyze in detail a wide range of scalar masses in the hundred GeV ballpark, paying particular attention to the transition as we increase this mass, from the merged $W/Z$ pair case (i.e. where the dedicated approach is needed) to them being well-separated (such that standard techniques suffice). 

An outline of the rest of this paper is as follows. We begin in section \ref{bound} with a discussion of the constraints on such particles coming from their direct production via VBF or via associated production. In this section, we discuss the above-mentioned cases of tensor and vector models separately. Section \ref{strategy} then develops the strategy for tagging a boosted scalar from its decay into a merged diboson using a toy model which is outlined at the beginning of the section. We have two possibilities for $W/Z$ decay here: fully hadronic and semi-leptonic, which are analyzed in subsections~\ref{hadronic_tagger} and~\ref{semileptonic_tagger} respectively. This is followed by the application of these strategies to the extended warped extra-dimensional framework in section~\ref{sec:application_to_warped_framework}. We first briefly review this scenario in subsection~\ref{warped_review}, and then present our results for the LHC signals for the boosted diboson resonance arising in it, in subsection \ref{warped_results}. We conclude in section \ref{conclude}.

\section{Constraints on Direct Production}
\label{bound}

As outlined in the introduction (see also Ref.~\cite{Agashe:2017wss} for more details), we can envisage two types of Lorentz structures in the gauge couplings of $\varphi$, dubbed ``tensor''  and ``vector'' respectively. The methodology advocated in the present study applies qualitatively to both vector and tensor cases. We will focus primarily on the tensor case, to be more concrete, and make some remarks on the vector case later to highlight the similarities and differences. For the tensor case, we discuss existing experimental searches that are sensitive to direct production of a light scalar, and present the allowed parameter space. Most of the discussions of this section will be based on a simplified model where the only BSM particle is a scalar $\varphi$ and it couples only to a pair of EW gauge bosons. The results of this section will demonstrate that current search methods leave open the possibility of light scalars, and motivates the exploration of alternative production topologies in which they might be discovered.

\subsection{Tensor model}

The tensor model is characterized by the scalar $\varphi$ coupling to the field strength tensors of gauge bosons. Assuming CP-even $\varphi$, the lowest dimensional operators are given by
\begin{align}
{\cal L}
& \ni
- \frac{1}{4 g_W^2} W_{\mu\nu}^2 - \frac{1}{4 g_B^2} B_{\mu\nu}^2 - A_W \frac{\varphi}{\Lambda} W_{\mu\nu}^2 - A_B \frac{\varphi}{\Lambda} B_{\mu\nu}^2 
\nonumber \\
& \to 
- \frac{1}{4} W_{\mu\nu}^2 - \frac{1}{4} B_{\mu\nu}^2 - A_W \frac{\varphi}{\Lambda} g_W^2 W_{\mu\nu}^2 - A_B \frac{\varphi}{\Lambda} g_B^2 B_{\mu\nu}^2 \nonumber \\
& = 
- \frac{1}{2} W^+_{\mu\nu} W^{- \mu\nu} - \frac{1}{4} Z_{\mu\nu}^2 - \frac{1}{4} \gamma_{\mu\nu}^2 
 \nonumber \\
& 
- \frac{\varphi}{\Lambda} 
\left[ 
2 A_W g_W^2 W^+_{\mu\nu} W^{- \mu\nu} 
+ \left( A_W g_W^2 c_W^2 + A_B g_B^2 s_W^2 \right) Z_{\mu\nu}^2  
\right. \nonumber \\
& 
+ \left. 
\left( A_W g_W^2 s_W^2 + A_B g_B^2 c_W^2 \right) 
\gamma_{\mu\nu}^2 
+ \left( A_W g_W^2 - A_B g_B^2 \right) 2 c_W s_W Z_{\mu\nu} \gamma^{\mu\nu} 
\right], 
\label{tensor1}
\end{align}
where $g_{W} (g_B)$ is the $SU(2)_L$ ($U(1)_Y$) SM gauge coupling and $W_{\mu\nu}$ and $B_{\mu\nu}$ are the respective field strength tensors.\footnote{One can also have a CP-odd coupling of the form $F\widetilde{F}$, e.g. for axion like particles. The phenomenology that we have studied will be very similar for this case.}
 Also, $A_{W/B}$ denotes Wilson coefficient of higher dimensional operator and $c_W$ ($s_W$) is the cosine (sine) of the weak mixing angle. As seen clearly from the first line in~\eqref{tensor1}, the above couplings between $\varphi$ and the SM EW gauge bosons are manifestly $SU(2)_L \times U(1)_Y$ invariant for any given $A_W$ and $A_B$. From the first to second line in~\eqref{tensor1}, we did field redefinition $V_\mu \to g_V V_\mu$ (where $V = W, B$) while the third to fifth lines are obtained by writing $W_\mu^{1,2,3}$ and $B_\mu$ in terms of mass eigenstates $W_\mu^\pm, Z_\mu$ and $\gamma_\mu$. We notice that for generic $A_W$ and $A_B$, the scalar couples to a $Z$ boson and a photon, in addition to its couplings to a pair of $W$, $Z$, and photon.

A familiar example of such a coupling can already be found within the SM: a SM top quark loop generates the couplings between SM Higgs and gluon/photon precisely in this form, with the suppression scale $\Lambda$ being the top mass. In our case, we have in mind TeV scale new physics inducing the couplings in Eq.~(\ref{tensor1}) between a new light scalar $\varphi$ and SM EW gauge bosons. So, we take $\Lambda$ to be a few TeV. For the sake of concreteness and simplicity, for our current study, we take $A_W = A_B$. With this Eq.~(\ref{tensor1}) becomes
\begin{align}
{\cal L} 
& \ni 
- 2 A_W \frac{\varphi}{\Lambda} \left[ g_W^2 W^+_{\mu\nu} W^{- \mu\nu} + \frac{1}{2} \left( g_W^2 c_W^2 + g_B^2 s_W^2 \right) Z_{\mu\nu}^2 + \frac{1}{2} \left( g_W^2 s_W^2 + g_B^2 c_W^2 \right) \gamma_{\mu\nu}^2 \right. \nonumber \\ 
& 
\left. \hspace{1.8cm} + \frac{{}}{{}} \left( g_W^2 - g_B^2 \right) c_W s_W Z_{\mu\nu} \gamma^{\mu\nu} \right] \nonumber \\
& = 
- \frac{\varphi}{\Lambda_{\rm eff}} \left[ g_W^2 W^+_{\mu\nu} W^{- \mu\nu} + \frac{1}{2} \left( g_W^2 c_W^2 + g_B^2 s_W^2 \right) Z_{\mu\nu}^2 + \frac{1}{2} \left( g_W^2 s_W^2 + g_B^2 c_W^2 \right) \gamma_{\mu\nu}^2 \right. \nonumber \\
& 
\left. \hspace{1.8cm} + \frac{{}}{{}} \left( g_W^2 - g_B^2 \right) c_W s_W Z_{\mu\nu} \gamma^{\mu\nu} \right], 
\label{eq:tensor_coupling_simplified_model}
\end{align}
where we have defined $\Lambda_{\rm eff} \equiv \Lambda / (2 A_W)$.
As already mentioned, the above interactions allow the direct production of the scalar $\varphi$; in particular, {\em single} production can occur via vector boson fusion (VBF), including $WW$, $ZZ$, $Z\gamma$ and $\gamma\gamma$ fusions. 
Moreover, $\varphi$ can also be produced in {\em association} with $W/Z/\gamma$ via $s$-channel $W/Z/\gamma$ exchange. The direct production rate can be parameterized by two parameters: $m_\varphi$ and $\Lambda_\text{eff}$. The VBF production rate is similar to or larger than associated production for the parameter ranges we consider, with $\sigma_\text{VBF} / \sigma_\text{AP}$ ranging between 0.7 and 5.3 for $m_\varphi$ between 200~GeV and 1~TeV respectively. For heavier masses such that phase space suppression can be neglected, the singlet scalar $\varphi$ decays to a pair of SM gauge bosons with the following hierarchy of branching ratios: 
\begin{align}
\textrm{BR}(\varphi\to WW) > \textrm{BR}(\varphi\to ZZ) > \textrm{BR}(\varphi\to \gamma\gamma) > \textrm{BR}(\varphi\to Z\gamma)\:.
\end{align}
However, when $m_\varphi$ starts to get small, e.g. around $200$ GeV, the $\textrm{BR}(\varphi\to ZZ)$ can be smaller than $\textrm{BR}(\varphi\to \gamma\gamma)$, due to phase space suppression.
Note that $\varphi$ produced in this manner is typically {\em not} boosted so that standard tagging can be applied to it once it decays into dibosons; each boson could be boosted or unboosted depending on the mass gap between $\varphi$ and the boson.

\begin{figure}[t]
\centering
\includegraphics[width=0.6\linewidth]{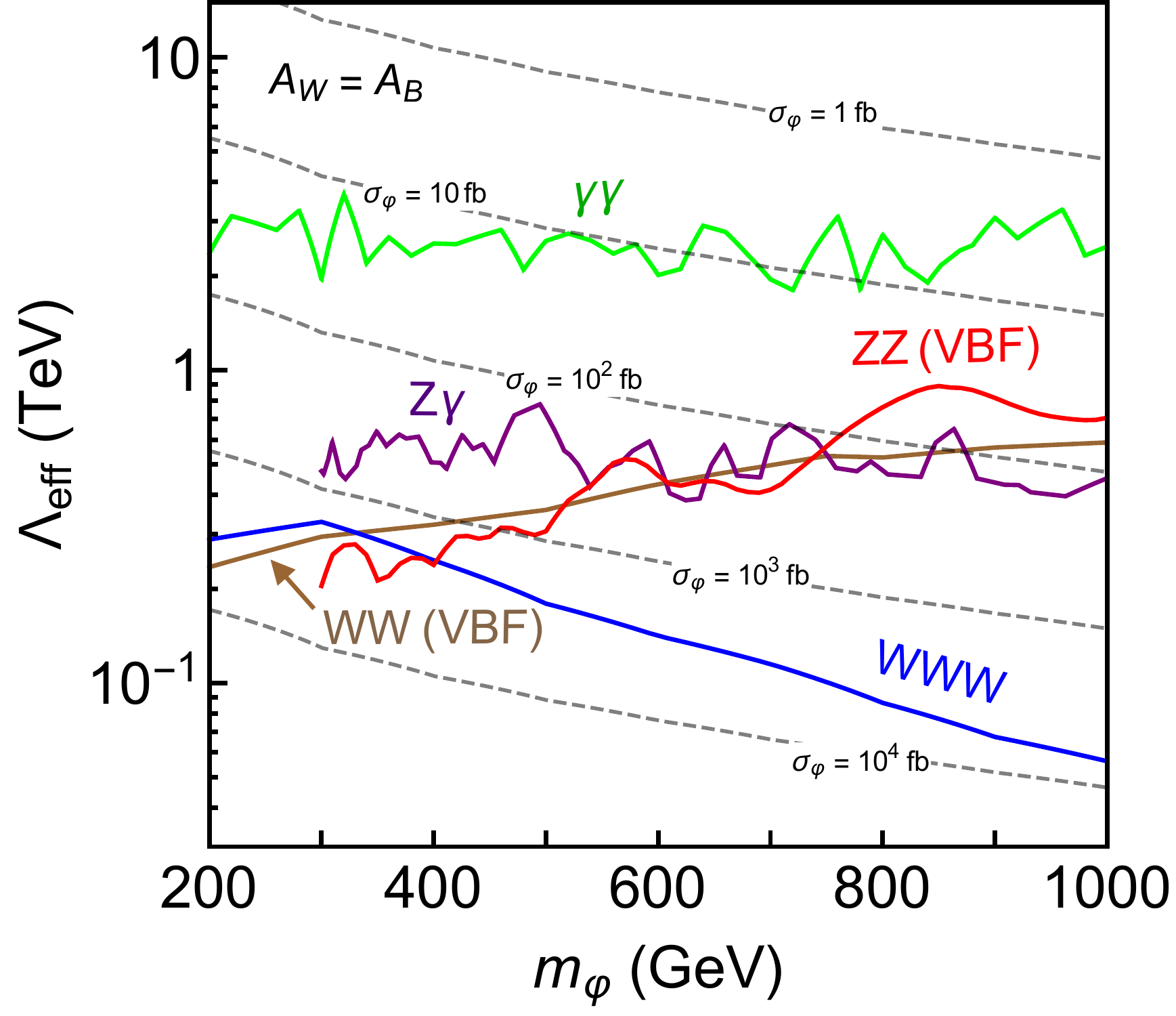}
\caption{
\small{Bounds on tensor-type coupling of scalar to SM EW gauge bosons in the $m_\varphi-\Lambda_{\rm eff}$ plane. Five relevant searches are shown: ATLAS diphoton (green), ATLAS diboson $WW$ (VBF, fully leptonic, in brown), ATLAS diboson $ZZ$ (VBF, $\ell \ell q q  + \nu \nu q q$ in red), CMS $Z \gamma$ (leptonic, in purple), and ATLAS triboson $WWW$ (blue). Regions below the solid lines are excluded by the corresponding experiments. The dashed lines show the scalar production cross section $\sigma(p p \to \varphi)$ (sum of both VBF and associated production). For $ZZ$ and $Z\gamma$, results are shown only down to 300 GeV because relevant experimental search results are reported only to that point.}
}
\label{fig:bound_tensor}
\end{figure}

There are several existing experimental search channels relevant to a light scalar produced via the coupling defined in Eq.~(\ref{eq:tensor_coupling_simplified_model}). The combined bounds in the $m_\varphi$-$\Lambda_{\rm eff}$ plane are shown in Fig.~\ref{fig:bound_tensor}, where we consider the following constraints.

\begin{itemize}
\item As the scalar may couple to a pair of photons, diphoton searches such as in Ref.~\cite{Aaboud:2017yyg} can provide very strong limits, mainly due to the cleanliness of the resulting final state. Indeed, we find that for the tensor model, the ATLAS diphoton resonance search imposes the most stringent bound on the parameter space (see Fig.~\ref{fig:bound_tensor}).

\item Its decays into pairs of $W$, $Z$ can be constrained by the standard diboson searches. For the $WW$ diboson decay, and for $m_\varphi \sim$ a few hundred GeV that we are interested in this paper, we find that $WW\to e\nu\mu\nu$ channel is the most efficient \cite{Aaboud:2017gsl}, compared to $\ell\nu j j $ \cite{Aaboud:2017fgj} or fully hadronic \cite{Aaboud:2017eta} channels, which are subject to larger backgrounds. For the $ZZ$ decay mode we consider the search of Ref.~\cite{Aaboud:2017itg} in the $\ell \ell q q$ and $\nu \nu q q$ final states. Because the limits in the VBF search channels are the strongest in these searches, and because the VBF production cross section in the model is similar to or larger than associated production, we use the limits that these searches placed on scalar resonances produced through VBF to constrain our model.

\item When $\varphi$ decays to a $Z$ and a photon, $Z\gamma$ resonance search can be relevant. In particular, Ref.~\cite{Sirunyan:2017hsb} performed analysis considering $Z$ decaying both leptonically (into electron or muon pair) and hadronically (forming one fat jet $J$). We use the limits placed by the leptonic channel of this search, which is more sensitive to the narrow resonance.

\item When the scalar is produced in association with $W$ or $Z$, the final state consists of three EW gauge bosons. Especially, the final state $W^{\pm} W^{\pm} W^{\mp}$ is constrained by triboson searches. Two decay channels, $W^{\pm} W^{\pm} W^{\mp} \to \ell^{\pm} \nu \ell^{\pm} \nu \ell^{\mp} \nu$ (fully leptonic) and $W^{\pm} W^{\pm} W^{\mp} \to \ell^{\pm} \nu \ell^{\pm} \nu jj$ (semi-leptonic), are investigated by the ATLAS Collaboration and limits are reported in Ref.~\cite{Aaboud:2016ftt} using $\sqrt{s} = 8 \; \text{TeV}$ data. 
Firstly, event rates are reported in six mutually exclusive search regions, depending on the flavors and signs of the final state leptons. Of these, we find that the category ($e^\pm e^\pm$) of $\ell \nu \ell \nu jj$-channel places the strongest limits for radion masses $m_\varphi \lesssim 600 \; \text{GeV}$. For a higher mass radion, the final state is relatively hard compared to the SM backgrounds, and stronger limits can be obtained using the distributions in $m^{3\ell}_T$ (the transverse mass of the three leptons in the $\ell \nu \ell \nu \ell \nu$ final state) and $\sum p_T$ (the sum over $p_T$ for the leptons, jet, and missing energy in the $\ell \nu \ell \nu j j$ final state), which were used in the ATLAS study to constrain BSM contributions to the gauge quartic coupling.
We obtain approximate limits that can be derived from these distributions using a simple approach. Above $m_T^{3 \ell} > 400 \; \text{GeV}$ and $\sum p_T > 700 \; \text{GeV}$ the channels are nearly background-free, and so a signal would be ruled out at the $2 \sigma$ confidence level if it predicts more than 3 events in these regions (since the probability for a Poisson fluctuation of three predicted signal events down to zero observed events is 0.05). Such an approximate limit will be conservative, in the sense that no parameter space will be incorrectly ruled out. While a stronger limit could be obtained by a full shape analysis, the strongest part of the constraint will come from these background-free regions anyway. In order to compute the signal rate in these regions, we simulate using \texttt{MadGraph@aMC}~\cite{Alwall:2014hca} for parton-level events generation (with center of mass energy matching that of the search), \texttt{Pythia6}~\cite{Sjostrand:2006za} for parton shower and hadronization, and  \texttt{Delphes3}~\cite{deFavereau:2013fsa} for detector simulation, and apply the cuts described in Ref.~\cite{Aaboud:2016ftt}. The $WWW$ bound shown in Fig.~\ref{fig:bound_tensor} is taken as the strongest of the bounds coming from the six signal regions and from the  $m_T^{3 \ell}$ and $\sum p_T$ distributions at each radion mass.

\end{itemize}
Taking all these constraints into account, current bound on $\Lambda_{\rm eff}$ from the scalar direct production in tensor model is $\Lambda_{\rm eff} \gtrsim 2-3$ TeV, primarily from the $\gamma\gamma$ constraints. 

We point out that the couplings in Eq.~(\ref{tensor1}) materialize in a suitable model within a general warped extra dimensional framework (for more details, see our earlier works \cite{Agashe:2016rle, Agashe:2016kfr, Agashe:2017wss}). The scale $\Lambda$ in Eq.~\eqref{eq:tensor_coupling_simplified_model} corresponds to the KK scale, at which the KK excitations of the EW gauge bosons appear.  
Of course, KK EW gauge bosons also couple to quarks/leptons, which then is relevant for their production and decays at the LHC. Namely, the couplings of KK EW gauge bosons to quarks enable their single production at the LHC and their couplings to leptons and quarks allow us to look for them in lepton and/or dijet final states. These searches constrain the KK scale to be a few TeV, implying suppressed direct production of $\varphi$, and allow for a light $\varphi$ consistent with the bounds in Fig.~\ref{fig:bound_tensor}. These KK EW bosons also have couplings to transverse $W/Z/\gamma$ and to $\varphi$, allowing their decay into these states and the possibility of discovering $\varphi$ in this topology.

\subsection{Vector model}

Instead of coupling to the field strength tensors of SM gauge fields, the singlet scalar $\varphi$ can also directly couple to a pair of SM gauge boson vector potentials (e.g. $\varphi W^{\mu}W_{\mu}$), just like the SM Higgs ($h$). We call the model having this type of couplings as ``vector model''. The simplest way to realize such a coupling structure may be to mix the scalar with the SM Higgs. Such a mixing can naturally arise in a model with a generic scalar potential consisting of $h$ and $\varphi$ through a Higgs portal coupling $|\varphi|^2 |H|^2$, or through a cubic term $\varphi |H|^2$ if allowed by the symmetries of the model. This type of vector models are often parametrized by two new parameters: $m_\varphi$ and $\sin\alpha$, where $\sin\alpha$ describes the mixing between $h$ and $\varphi$. There are a lot of studies related to the current constraints of such models, and generically $\sin^2\alpha$ is constrained to be less than 0.1 (see, for example, \cite{Robens:2016xkb}).

While we are not focusing on this case explicitly, there are some similarities and differences to the tensor case that can be pointed out from a phenomenological point of view. The key similarity is that a scalar whose decays are due entirely to their mixing with the Higgs will decay dominantly into pairs of vector bosons, so long as these are kinematically available. Di-Higgs decays might compete depending on the couplings of the model, and $t\bar{t}$ decays will be subdominant even above threshold, due to the longitudinal enhancement for the vector boson decays. This is a generic feature of scalars in this mass range whose decays are determined by Higgs portal type couplings, as might be the case for example for the lightest state in a hidden sector. The main phenomenological difference is the much smaller branching fraction into diphotons, which severely weakens the direct search sensitivity compared to the tensor model. On the other hand, the main production channel will be via gluon fusion which gives a parton distribution function enhancement for the production cross section. Nonetheless, the direct constraints on this model are weak above 300 GeV, and the dominant constraints are coming from precision $W$ mass measurements~\cite{Robens:2016xkb}.

In more sophisticated vector models, new scalars could be associated with extensions of the EW {\em gauge} symmetry, for example, left-right symmetric (i.e. $SU(2)_L \times SU(2)_R \times U(1)_{ B - L }$) models \cite{Mohapatra:2016twe}. Such models then also feature heavier gauge bosons, a.k.a. $W^{ \prime }/Z^{ \prime }$, with couplings to $\varphi$ like $\varphi W^{\mu}W'_{\mu}$. As in the tensor model (i.e. like KK $W/Z$ gauge bosons), $W^{ \prime }/Z^{ \prime }$ couple to quarks/leptons as well. One can then produce $W^{ \prime }/Z^{ \prime }$ via these couplings and it may subsequently decay into a boosted $\varphi$ through the $\varphi W^{\mu}W'_{\mu}$/$\varphi Z^{\mu}Z'_{\mu}$ coupling. Just as for the tensor model, a boosted $\varphi$ decaying to $WW$ or $ZZ$ becomes an interesting signature, which will be the focus of the following section.

\section{Boosted $WW$ Taggers}
\label{strategy}

\begin{figure}[ht]
\centering
\begin{subfigure}[b]{0.45\textwidth}
\includegraphics[width=\textwidth]{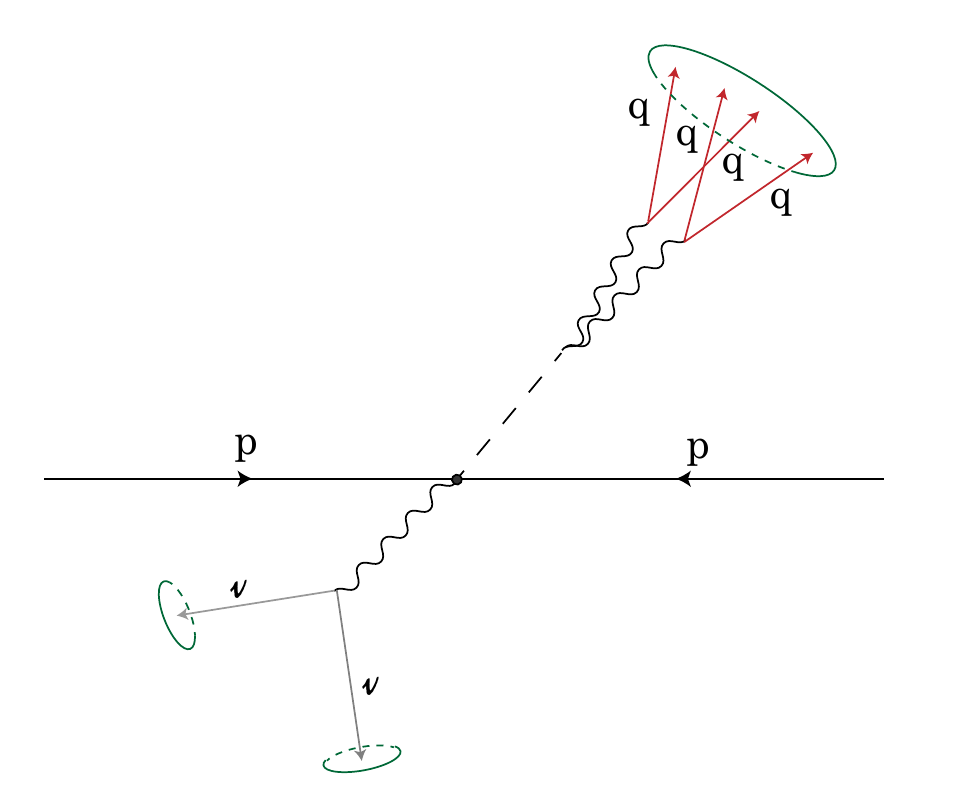}
\end{subfigure}
\begin{subfigure}[b]{0.45\textwidth}
\includegraphics[width=\textwidth]{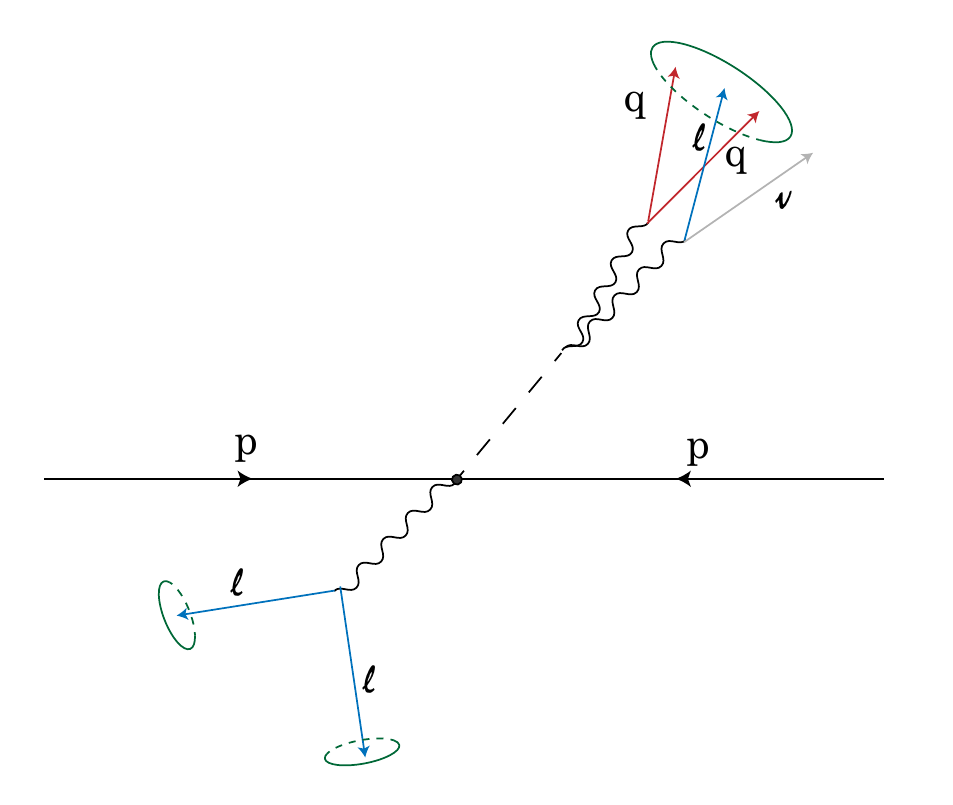}
\end{subfigure}
\caption{
\small{Considered signal processes that motivate dedicated tagging strategies. Left: a singly-produced $Z'$ decays to an invisible $Z$ and a scalar which decays further to a pair of fully hadronic $W$'s. Right: a singly-produced $Z'$ decays to a leptonic $Z$ and a scalar which decays further to a pair of $W$'s, one of which decays leptonically and the other hadronically.}
}
\label{fig:LeptonInJet}
\end{figure}

In this section, we describe the methods used to tag a boosted diboson jet -- a collimated pair of $W$ or $Z$ bosons falling within a single jet. We will focus on the decay into $WW$, though the $ZZ$ decay will have many similarities. There are two cases of interest depending on the decay mode of the $W$'s from the $\varphi$ decay -- either both the $W$ decay hadronically, or one of them decays leptonically (see Fig.~\ref{fig:LeptonInJet}). 

Since we are interested in low masses for $\varphi$, the $W$'s from its decay, \textbf{and} the decay products of the $W$'s can gets merged, depending on the mass and $p_T$ of $\varphi$. It is instructive to compare the typical $\Delta R$ separations, to be able to see clearly the motivation for the various approaches we take for the tagging, both for the hadronic and the semi-leptonic cases. 

There are two angular scales relevant for the process $\varphi \to WW \to qqqq/qq\ell\nu$. Firstly, for $p_{T,\varphi} \gg m_\varphi$ the two $W$ bosons in the jet will have a typical angular scale given by
\begin{align}
\Delta R_{WW} \simeq \frac{2 \sqrt{m_\varphi^2 - 4 m_W^2}}{p_{T, \varphi}}.
\label{eq:dr-ww}
\end{align}
Next, the $qq/\ell\nu$ from a $W$ decay will have a typical angular separation
\begin{align}
\Delta R_{qq/\ell\nu} \simeq \frac{2 m_W}{p_{T, W}} \simeq \frac{4 m_W}{p_{T, \varphi}},
\label{eq:dr-qq}
\end{align}
where in the last step we have assumed that the transverse momentum of $\varphi$ is almost equally distributed to each $W$'s. In the limit of $m_{\varphi} \gg 2 m_W$ so that we can drop the $m_W$ term in Eq.~\eqref{eq:dr-ww}, we see a distinct hierarchy of two-prong substructures at these two angular scales corresponding to the larger $\varphi \to WW$ splitting and the smaller $W \to qq/\ell\nu$ splittings, that is, 
\begin{align}
\Delta R_{WW} \simeq \frac{2m_\varphi}{p_{T,\varphi}} \quad > \quad   \Delta R_{qq/\ell\nu} \simeq \frac{4m_W}{p_{T,\varphi}}\:. 
\label{eq:dr-comparision}
\end{align}
For example, $p_{T, \varphi} = 1.5 \; \text{TeV}$ and $m_\varphi = 500 \; \text{GeV}$ give rise to $\Delta R_{WW} \simeq 0.63$ and $\Delta R_{qq/\ell\nu} \simeq 0.22$. 

However, as $m_\varphi$ starts getting close to $2m_W$, the result in~\eqref{eq:dr-comparision} is not correct anymore and this hierarchy in $\Delta R$ is lost. For example, comparing Eq.~\eqref{eq:dr-ww} to Eq.~\eqref{eq:dr-qq} shows that if $m_\varphi \lesssim \sqrt{20} \, m_W \simeq 360$ GeV, the angular separation $\Delta R_{WW} \lesssim 2 \Delta R_{qq/\ell\nu}$ for all  $p_{T,\varphi} \gg m_\varphi$, and the decay products from the decays of the two $W$'s begin to overlap. Hence it is not easy to separate the two $W$ bosons.

So far, the discussion holds independent of whether we consider the fully hadronic mode or the semi-leptonic mode. However an important distinction arises when one wants to identify the individual decay products from the two $W$'s, and therefore the $W$'s themselves. For the semi-leptonic case, if one can identify the lepton from the $W$ then that is sufficient to reconstruct the two $W$'s. In this case, the ``diboson-jet'' is expected to have two hadronic activity centers in addition to the lepton and they must be from the hadronic $W$ decay, independent of whether the condition in~\eqref{eq:dr-comparision} is satisfied or not. However, in the fully hadronic case, the diboson-jet has four centers of hadronic activities, and there is no unique way to associate them to the two $W$'s, when the hierarchy of ~\eqref{eq:dr-comparision} does not hold, i.e. for lighter $\varphi$ masses. This observation motivates us to use two different boosted-diboson tagging strategies for the fully hadronic case, to target the two different regimes, while one tagging strategy suffices for the semi-leptonic case. We describe these strategies in the next subsections. 

\begin{figure}[ht]
\centering
\includegraphics[width=0.65\textwidth]{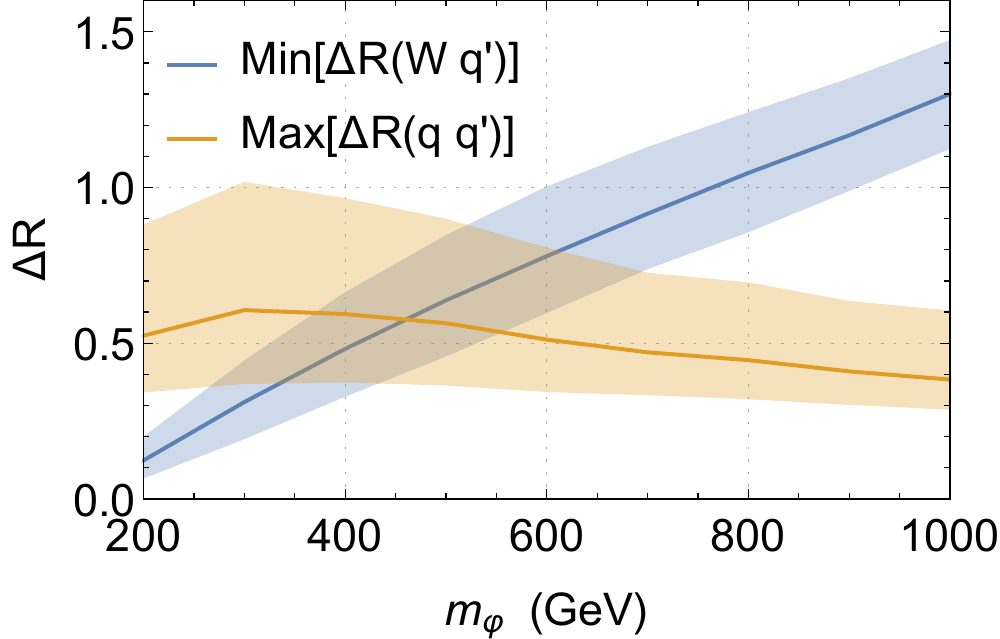}
\caption{
\small{Parton-level separations within a boosted diboson jet produced from the decay of a 3~TeV parent resonance. Orange: in each event, the larger of the two angular separations between two quarks from a $W \to q \bar{q}$ decay. Blue: in each event, the smallest angle between a $W$ boson and a quark coming from the other $W$ boson. In each case, the solid line represents the mean over the event sample, while the shaded band represents the range between the 16th and 84th percentile (one sigma).}
}
\label{fig:dr-WW-vs-dr-jj}
\end{figure}

In order to verify the conclusions drawn from the above approximations, we study a parton level simulation of a toy model $Z' \to Z \varphi$, with $\varphi \to WW \to q\bar{q}q\bar{q}$ (this $Z'$ is a neutral KK resonance of the warped model described in Ref.~\cite{Agashe:2017wss}, which couples to the transverse components of the SM vector bosons). The $Z'$ mass is set to 3 TeV. In the case that all quarks are closer to their ``sibling'' quark from the same parent $W$ decay than they are to the opposite $W$, the jet exhibits a clear hierarchical two-prong substructure of two-prong substructures and the two $W$ bosons can be separated even in the case of the fully hadronic decay. In the case that a quark is closer to the opposite $W$ than to its sibling however, it will be difficult to resolve the two $W$ bosons. In Fig.~\ref{fig:dr-WW-vs-dr-jj} we have plotted for a range of $\varphi$ masses the larger separation between quark siblings in orange and the smallest separation between quark and opposite $W$ in blue (together with the 1-sigma error bands), after selecting only those jets in which all four quarks fall within 1.2 radians of the scalar momentum. At low masses, there is always a quark closer to the opposite $W$ than to its sibling, but by 400 GeV the diboson jets frequently exhibit well separated $W$'s and by 700 GeV the separation is very pronounced.

Finally, a comment needs to be made about using an invariant mass window cut on the scalar, when implementing the taggers discussed here. It is desirable to keep the tagger as generic as possible, so that its performance can be understood without too much dependence on the other details of the process to which it is applied. For the semi-leptonic case, $\varphi \to WW \to l \nu q q$, the ability to reconstruct the neutrino momentum, and therefore the scalar invariant mass requires no other missing energy in the process. That is certainly not generic, and therefore in this section we do not use any requirement that the mass of the partially leptonically decaying scalar can be reconstructed. There is no such complication in the fully hadronic case. Hence we use an invariant mass cut on the scalar only for the hadronic taggers discussed below. 

\subsection{Fully hadronic}
\label{hadronic_tagger}

We consider the case when both $W$'s from $\varphi$ decay hadronically (see the left panel of Fig.~\ref{fig:LeptonInJet}). We discuss here the two different tagging strategies to target the two different regimes -- depending on whether~\eqref{eq:dr-comparision} is satisfied or not, which in turn depends on $m_\varphi$.

In both cases, we begin by clustering a jet using the anti-$k_t$ algorithm~\cite{Cacciari:2008gp} with jet radius $R = 1.2$. We then recluster the jet constituents using the Cambridge-Aachen algorithm~\cite{Dokshitzer:1997in,Wobisch:1998wt} in order to obtain an angular-ordered cluster sequence. Both algorithms make use of the soft-drop grooming and tagging algorithm \cite{Larkoski:2014wba}. The soft drop algorithm takes a clustered jet, and considers the two constituents which are merged in the final step of the clustering sequence. The softer of these two constituents is discarded from the jet unless the soft drop condition is satisfied
\begin{align}
\text{Soft Drop Condition: } \frac{\text{min}(p_{T1}, p_{T2})}{p_{T1} + p_{T2}} > z_\text{cut} \left(\frac{\Delta R_{12}}{R_0}\right)^\beta,
\end{align}
where $p_{T1}, p_{T2}$ are the transverse momenta of the jet constituents, $R_0$ is the radius parameter for the original clustering algorithm, $\Delta R_{12}$ is the angular separation between the two jet constituent four-momenta, and $z_\text{cut}$ and $\beta$ are parameters of the algorithm. 
In the case that the soft drop condition is not satisfied, the softer constituent is discarded from the jet, and the procedure is iterated again on the remaining constituent. Once the soft drop condition is satisfied then a hard splitting has been found. Both constituents are kept and their combination is the resultant groomed jet.

The soft drop algorithm can be used in tagging mode, in which case a jet is rejected if no hard splitting is found.
 In order to find four-prong substructure, we apply the algorithm recursively to two branches selected by an initial application of the soft drop algorithm. At this stage, the choice of $R_0$ becomes ambiguous since these subjets do not have the same angular scale as the initial jet. Our prescription is to use $2 m_J / p_{T,J}$ in place of $R_0$ where this quantity is calculated at the start of each soft-drop recursion. In the first recursion, this quantity is $2 m_J / p_{T,J} \sim 2 m_\varphi / p_{T, \varphi}$, while in the second recursions it is given by $2 m_{J} / p_{T,J} \sim 2 m_W / p_{T, W}$ for each subjet.

This algorithm is a specific application of recursive soft drop, described in Ref.~\cite{Dreyer:2018tjj} which was published during the later stages of our study and which focuses on the application to jets resulting from boosted SM objects.

\subsubsection*{Case 0: fully resolved $W$'s}
For $m_\varphi \gtrsim p_{T, \varphi} / 2$, the two $W$ bosons from the $\varphi$ decay will be sufficiently resolved that they could be reconstructed separately as a pair of fat jets and $W$-tagged in the normal way. In this case we cluster jets using a radius parameter $R_0$, and tag them for two-prong substructure using the original soft drop algorithm. We require each candidate $W$-jet to fall in the range $65 \; \text{GeV} < m_{W, \text{cand}} \leq 95 \; \text{GeV}$.

\subsubsection*{Case 1: partially resolved $W$'s (intermediate tagger)}
For  $ p_{T, \varphi} / 2 > m_\varphi \gtrsim 360 \; \text{GeV}$, the two $W$ bosons from the $\varphi$ decay will usually be reconstructed as a single ``fat'' jet but are frequently reasonably well separated into a pair of two-prong subjets. We therefore apply the soft-drop tagger to find the first hard splitting within the jet, which we assume to be the $\varphi \to WW$ splitting. We then apply the soft drop tagger again on each of the two resulting subjets, assigning the subsequent
splittings to the $W \to qq$ decays. Any jets which fail this soft drop selection are rejected. We then require that the two selected $W$-candidates have masses satisfying $60 \; \text{GeV} < m_{W, \text{cand}} \leq 100 \; \text{GeV}$.

\begin{figure}[ht]
\centering
\begin{subfigure}[b]{0.3\textwidth}
\includegraphics[width=\textwidth]{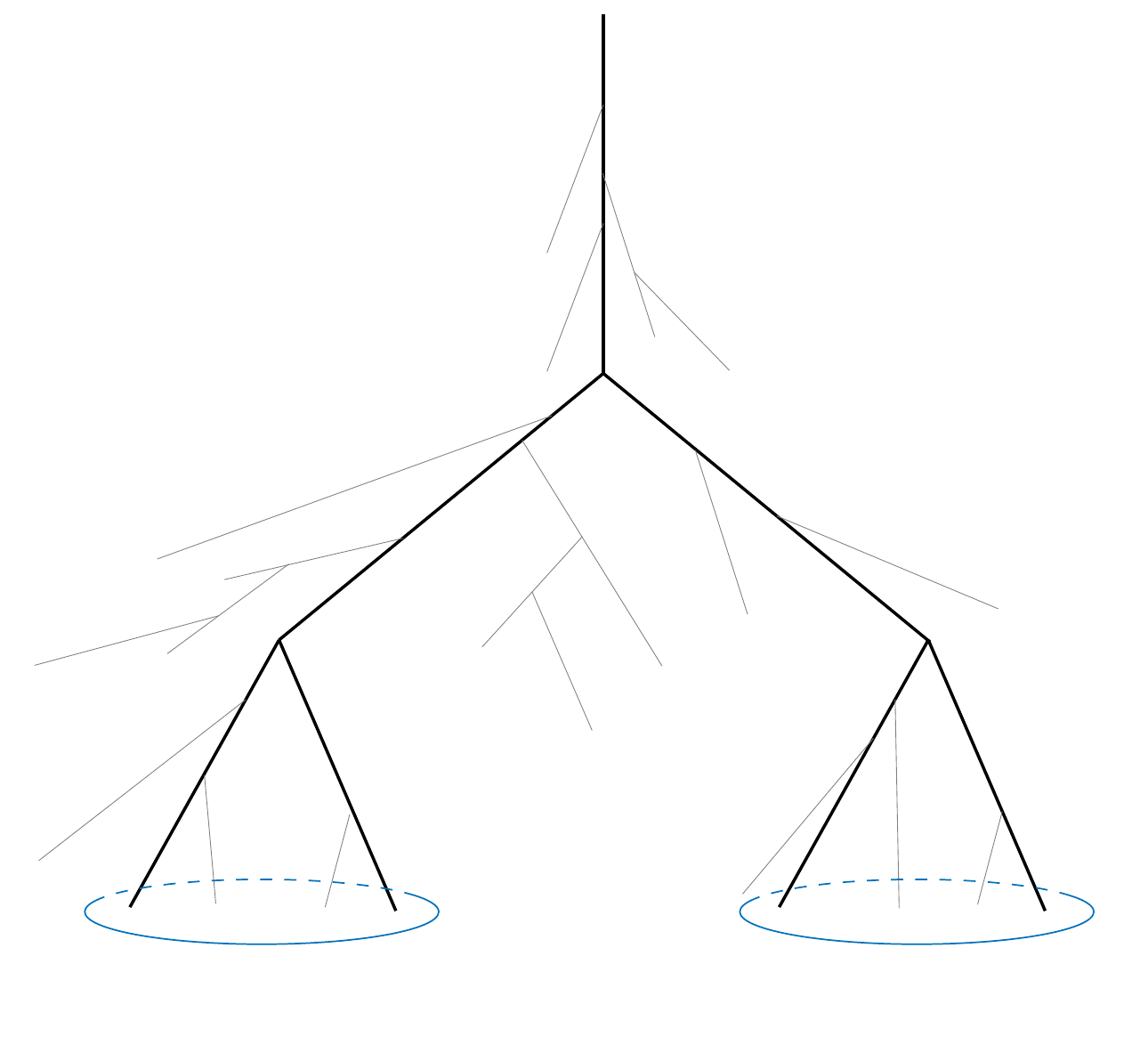}
\caption{Intermediate tagger}
\end{subfigure}
\begin{subfigure}[b]{0.2\textwidth}
\end{subfigure}
\begin{subfigure}[b]{0.6\textwidth}
\includegraphics[width=0.45\textwidth]{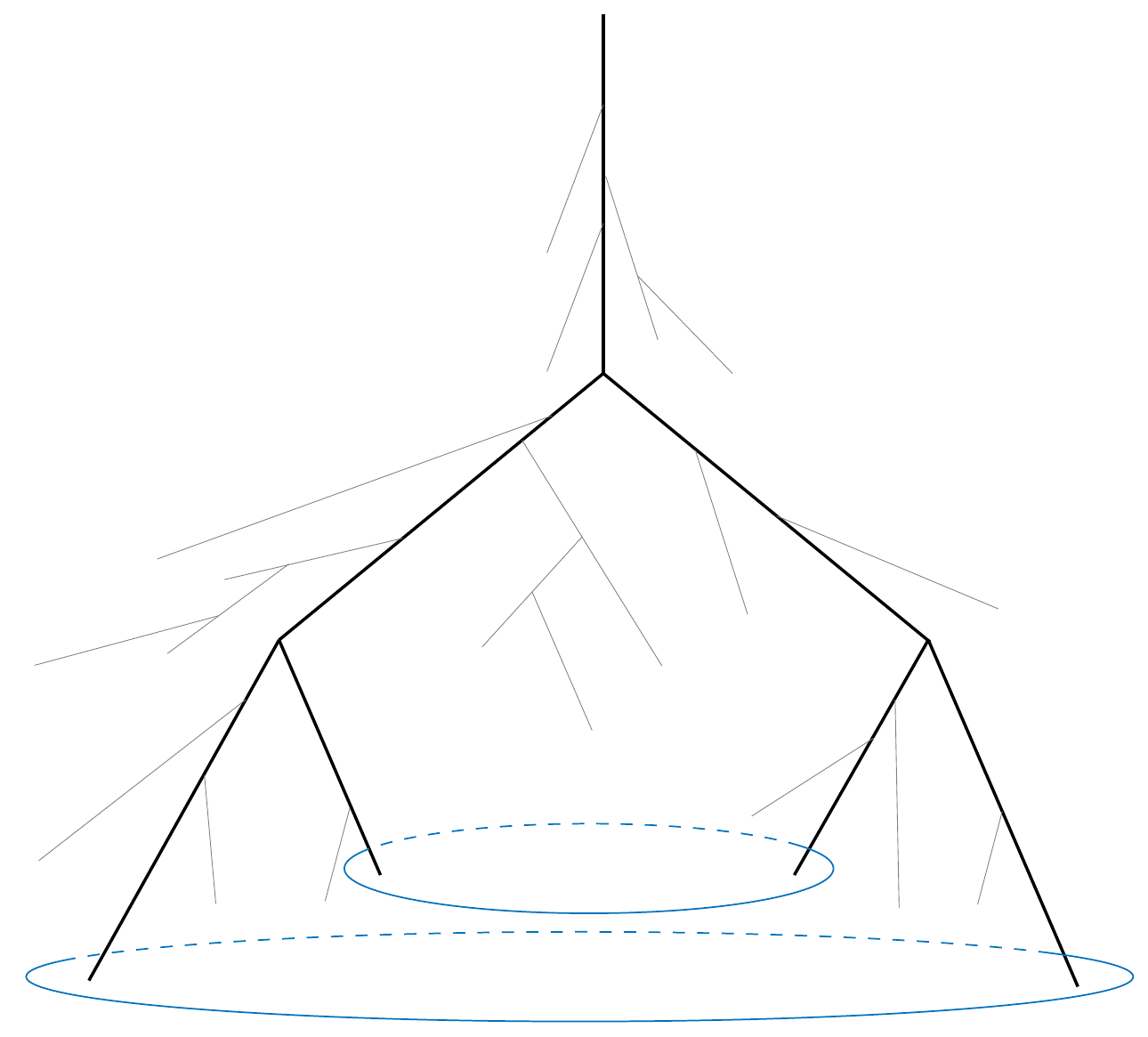}
\includegraphics[width=0.5\textwidth]{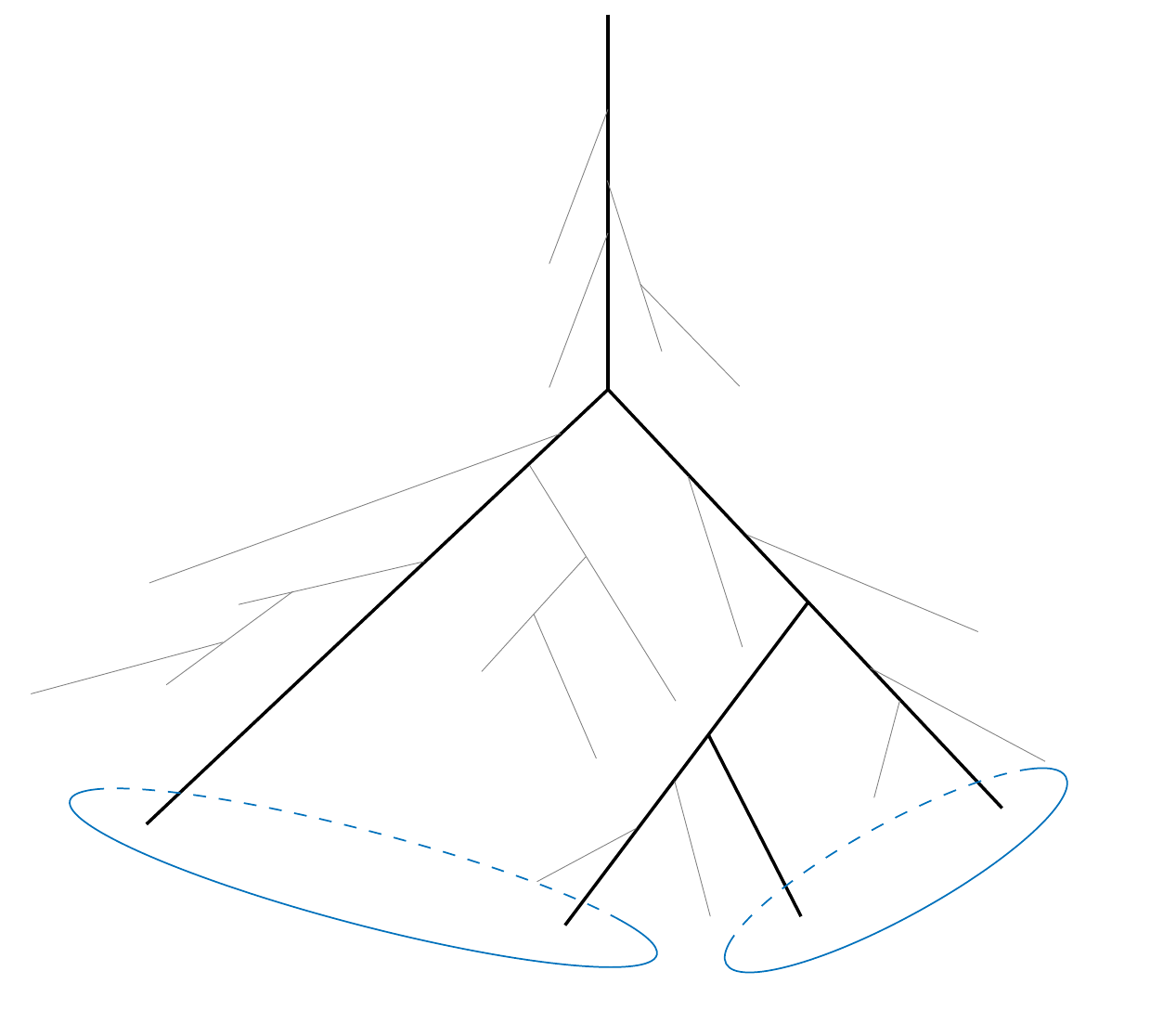}
\caption{Merged tagger}
\label{fig:merged_tagger_fig}
\end{subfigure}
\caption{
\small{Left: hard splitting tree (solid black lines) and assignment of subjets to $W$ candidates (blue ellipses) which is assumed by the intermediate tagger. Right: the two possible hard splitting trees allowed by the merged tagger. Grey lines show soft splittings. Also shown are example assignments of subjet pairs to $W$ candidates. These pairings are chosen to minimize the mass difference between the two $W$ candidates in the jet.}
}
\end{figure}

\subsubsection*{Case 2: fully merged $W$'s (merged tagger)}
In this case, for $m_\varphi \lesssim 360 \; \text{GeV}$, the two $W$ bosons from the $\varphi$ are fully merged and cannot be separated geometrically. We follow an approach similar to Ref.~\cite{Son:2012mb}. 
We seek four subjets by applying a soft drop tagger in multiple steps, but unlike case 1 we do not assume that the shape of the cluster sequence matches the shape of the decay chain. We go through all splittings in the jet, classifying each one as hard or soft according to the soft drop criterion. Of all the hard splittings, we choose the three with the largest parent jet mass. This constructs a splitting tree with three splittings, two internal lines, and four final daughters (which are not parents of any of these selected massive splittings). There are two possible tree structures, which are illustrated in Fig.~\ref{fig:merged_tagger_fig}. The four daughters are taken to be the quark-candidates. We then consider all possible assignments of quark candidate pairs into $W$-candidates, and select the assignment which gives the smallest difference between $W$-candidate masses. We then require that each $W$-candidate has a jet mass in the window $60 \; \text{GeV} < m_{W, \text{cand}} \leq 100 \; \text{GeV}$.

\subsubsection*{Performance}

In order to investigate the performance of these taggers, we conduct a Monte Carlo simulation study. We employ \texttt{MadGraph@aMC}~\cite{Alwall:2014hca} to generate parton-level events at LHC-14 TeV which are further pipelined to \texttt{Pythia8}~\cite{Sjostrand:2007gs} and then \texttt{Delphes3}~\cite{deFavereau:2013fsa} whose outputs are analyzed by our taggers.
For our signal sample, we consider a toy process $pp \to Z' \to Z \varphi$ followed by $\varphi \to WW \to qqqq$. We choose the mass of $Z'$ to be 3 TeV and let the prompt $Z$ decay to a pair of neutrinos to obtain a pure sample of boosted $WW$-jets.  The dominant background for fully hadronic diboson jets will be SM QCD jets. In order to obtain a sample of QCD jets in a similar topology to the signal we generate $Z + \text{jets}$, again with an invisibly decaying $Z$. In this case, we perform matching of 1- and 2-jet samples using the shower-kt scheme~\cite{Alwall:2008qv}. We generate signals for $\varphi$ in the range 200~GeV to 1000~GeV in order to study the transition between the merged and intermediate signal regimes, and also the transition to the fully-resolved $WW$ signal for higher $m_\varphi$. For the boosted taggers (cases 1 and 2), we cluster the Delphes particle flow output into jets using the anti-kt algorithm with jet radius $R = 1.2$. We consider only jets with $|\eta| < 2$ and take the hardest such jet as our diboson jet candidate, requiring it to have $p_T > 1 \; \text{TeV}$. Selected jets are then passed through the merged and partially resolved tagging algorithms which are described above. We performed a scan over parameters $\beta, z_\text{cut}$, and will present results using $\beta = -1.5$, $z_\text{cut} = 0.04$ which was found to give the best performance. Selected jets are finally required to have an invariant mass within 12.5\% of the signal hypothesis mass.

We also tested a strategy of attempting to reconstruct two $W$-jets as independent fat jets using standard $W$-tagging techniques (case 0). For this, we clustered jets using a $R = 0.8$ in one case and $R = 0.5$ in another. We again require all jets to satisfy $|\eta| < 2$ and select the hardest two such jets as the diboson candidate, requiring that the diboson candidate have $p_T > 1 \; \text{TeV}$ as before. The tagging criterion is as described in case 0: each jet should be tagged for two-prong substructure using soft drop in tagging mode and with parameters $\beta = 0, \; z_\text{cut} = 0.04$ (again, we performed a scan over these parameters and found this choice to be optimal), and each $W$ mass is required to be in the range $65 \; \text{GeV} < m_{W, \text{cand}} \leq 95 \; \text{GeV}$. The diboson candidate is also required to have an invariant mass within 12.5\% of the signal hypothesis mass.

\begin{figure}[t]
\centering
\begin{subfigure}[b]{0.45\textwidth}
\includegraphics[width=\textwidth]{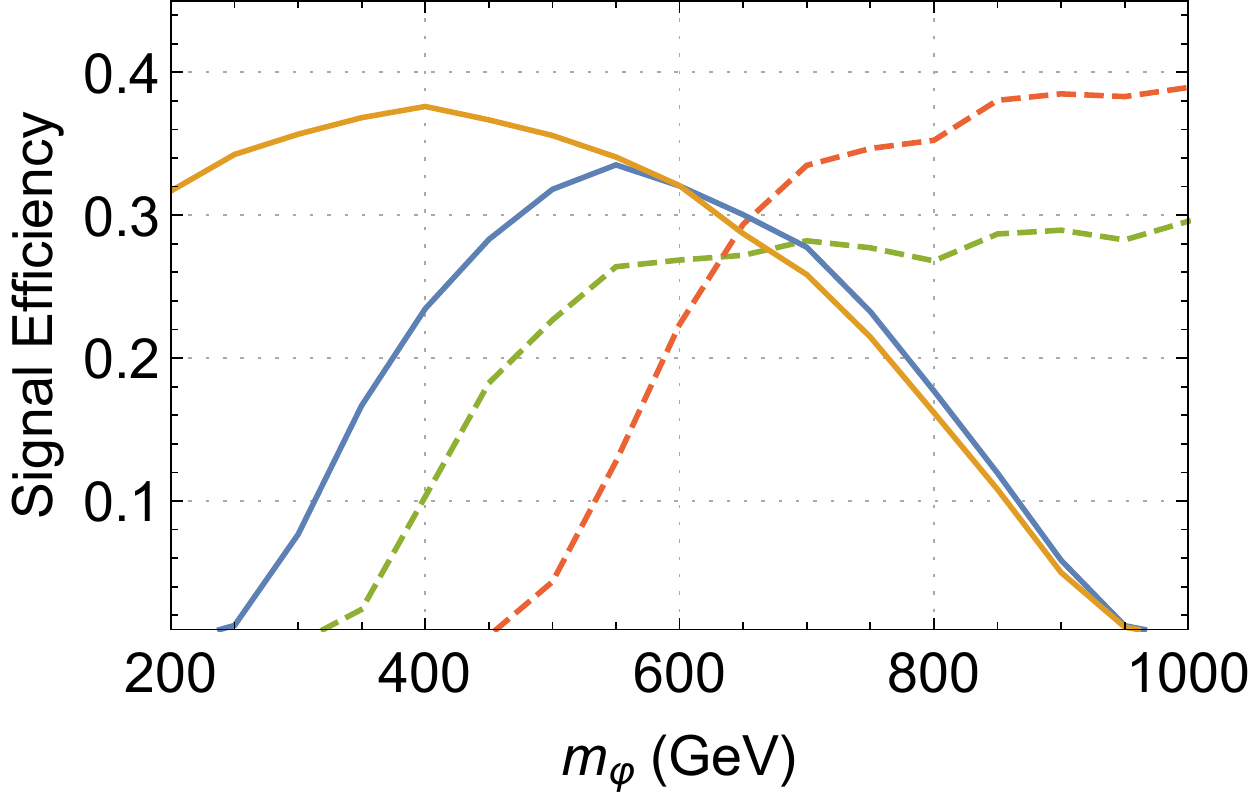}
\end{subfigure}
\begin{subfigure}[b]{0.45\textwidth}
\includegraphics[width=\textwidth]{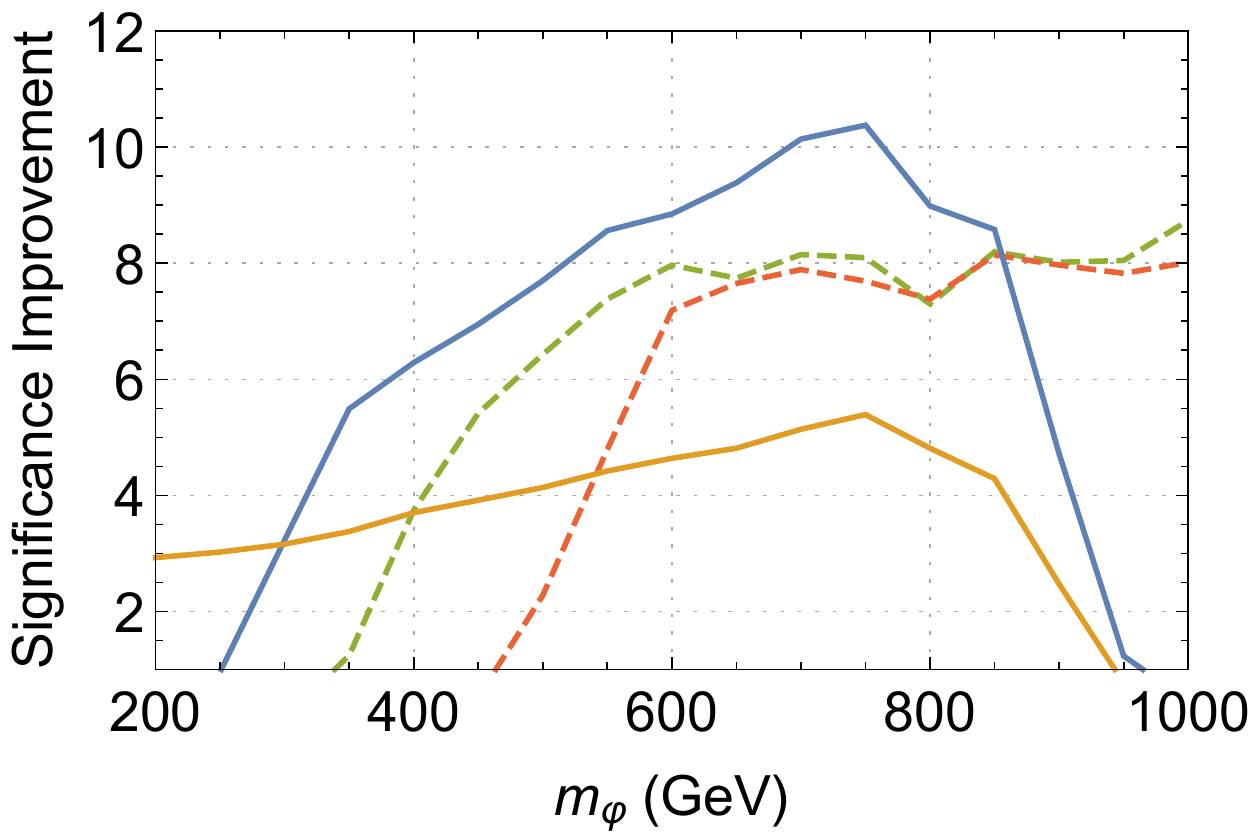}
\end{subfigure}
\caption{
\small{Tagger performance in terms of signal efficiency (left panel) and significance improvement (right panel) as a function of $m_\varphi$.  Lines are color-coded in the following way: solid orange for Merged tagger (case 2), solid blue for Intermediate tagger (case 1), dashed red/green for Resolved $W$-taggers (case 0) with $R = 0.5$ (green) and $R = 0.8$ (red).}
}
\label{fig:hadronic_massscan}
\end{figure}

In all cases, of those events passing the kinematic requirements, the fraction which additionally pass the tagging requirements and the diboson mass cut will be reported as the tagging efficiency for that strategy. The performance of the taggers are illustrated in Fig.~\ref{fig:hadronic_massscan} as a function of $m_\varphi$. The left panel shows signal efficiencies, while the right panel shows significance improvement defined by $\textrm{(signal efficiency)}/\sqrt{\textrm{background efficiency}}$. The orange and blue solid lines correspond to the merged and intermediate taggers respectively. The dashed lines correspond to fully resolved taggers (i.e. case 0), with green and red corresponding to using jet radius $R = 0.5, 0.8$ respectively.

Focusing first on the intermediate tagger, we find that indeed its signal efficiency drops off below around 350~GeV as the pair of $W$ bosons become completely merged, as was expected in our rough assessment before. The merged tagger shows high efficiency in the fully merged regime for which it was designed, but at the same time we observe that it competes in signal efficiency with the intermediate tagger even for higher masses of 500~GeV or more. On the other hand, its ability to reject background is somewhat poor in this regime, which leads to its significance improvement curve being inferior to that for the intermediate tagger for masses above 300~GeV. This is because the merged tagger considers all combinations of $W$-candidate assignments, giving a QCD jet many opportunities to find a mass-balanced pair of subjets. Secondly, the resolved algorithms with $R=0.8$ and $R=0.5$  have good discrimination performance down to $\varphi$ masses of 600~GeV and 450~GeV, respectively. In summary, we find that the merged and intermediate diboson taggers provide good coverage for boosted diboson signals with $m_\varphi \lesssim p_{T, \varphi} / 2$, while traditional resolved techniques provide good performance for higher masses.

\subsection{Semi-leptonic}
\label{semileptonic_tagger}
We next consider the case when one of the $W$ from $\varphi$ decays leptonically, while the other decays hadronically. As elaborated earlier, for low mass of $\varphi$, the resulting decay products get merged. In particular we expect a three-pronged jet, with one of the prongs coming from the energetic lepton inside (see the right panel of Fig.~\ref{fig:LeptonInJet}). Standard lepton isolation criteria will not suffice to construct the 4-momentum of the lepton, as they require no hard objects within a certain angular distance from the lepton.
Being able to isolate such a lepton is necessary if one wants to construct the corresponding neutrino 4-momentum, and therefore the invariant mass of the scalar. For this reason, simply treating the diboson-jet as a fat jet and using suitable substructure variables to tag it are not enough, since that would not treat the lepton inside the jet any different from other hadronic activity centers. 

It is possible to use suitably modified substructure techniques to identify the non-isolated energetic lepton inside the jet. Such identification methods have been developed in the past, in the context of semi-leptonic decays of boosted Higgs~\cite{Son:2012mb} and R-parity violating supersymmetry~\cite{Brust:2014gia}. The key observation is that when there is an energetic lepton buried inside a fat jet, it typically carries the biggest fraction of the $p_T$ of the subjet with which it is associated. This motivates defining a substructure based variable, lepton subjet fraction~\cite{Brust:2014gia}, denoted by $\LSF{n}$. First the jet is reclustered into $n$ exclusive subjets, and then $\LSF{n}$ is defined as follows
\begin{align}
\LSF{n} &= \:\:\underset{\text{\scriptsize{all leptons}}}{\text{max}}\:\:\frac{p_{T_{\ell_k}}}{p_{T_{s_j}}}\:, 
\end{align}
where $p_{T_{\ell_k}}$ is the transverse momentum of the $k^{\text{th}}$ lepton in subjet $s_j$, $p_{T_{s_j}}$ is the transverse momentum of the subjet $s_j$, and $n$ is the number of subjets considered, $1\leq j \leq n$. It is clear that $\LSF{n}$ takes values between 0 and 1, and signal-like events will tend to have a higher value of $\LSF{n}$. 

For the semi-leptonic diboson-jet we choose $n=3$, since the fat jet is expected to be three-pronged from the knowledge of the underlying parton level process. If the scalar is light, there will be an energetic lepton inside the fat-jet, resulting in a higher value of $\LSF{3}$, i.e. a value close to unity. Ordinary QCD backgrounds will tend to have low values for $\LSF{3}$, leaving only backgrounds involving the production of hard leptons such as $t\bar{t}$ and $W$ emission.

To demonstrate the discriminating power of this variable, 
we generate a sample of boosted $\varphi$ particles from the decay of a 3~TeV $Z'$, $p p \to Z' \to Z \varphi$ using \texttt{MadGraph5@aMC}. We consider the decay $\varphi \to WW \to qq \ell \nu$, with $\ell = e,\mu$, and choose the decay $Z \to \ell \bar{\ell}$. The overall process is shown schematically in the right panel of Fig.~\ref{fig:LeptonInJet}. Events are showered with \texttt{Pythia6} and detector response is simulated with \texttt{Delphes3}. We require events to contain a same-flavor opposite-sign lepton pair consistent with coming from the $Z$ boson decay, with an invariant mass in the window $(70, 110)$ GeV. After removing these leptons from the event, the remaining hard activity is resulting from the scalar decay products.

There are various standard model objects which might plausibly fake a semileptonic diboson jet. Firstly, normal QCD jets will frequently contain bottom, charm, or strange quarks whose decays may occasionally result in a hard lepton embedded inside the jet. Secondly, leptonic decays of boosted top quarks result in a hard lepton which might be embedded within the top jet. Finally, a very hard quark jet might emit a $W$ boson in the forward direction ($W$-strahlung), again resulting in an embedded lepton. We wish to generate high $p_T$ samples of these objects in a similar topology as the sample of boosted $\varphi$ particles in order to estimate the performance of the tagger in a fair way.

We generate samples of high-$p_T$ QCD jets recoiling against a leptonically decaying $Z$ boson, as for the $\varphi$ sample. We generate a separate $Zj$ and $Zb$ sample for light and $b$ jets respectively, and the QCD jet is identified as the hardest jet in the event after a leptonic $Z$ boson has been identified and removed. Top jets are obtained from a leptonic $t\bar{t}$ sample with the hardest jet in the event selected, while $W$-strahlung jets are obtained from $jjW$ production, again with the hardest jet in the event selected.

After generating detector-level samples for the signal and background objects, jets are constructed from the track and tower hits with no overlap with the leptons identified from the $Z$ decay, $\ell^\pm_z$ (where relevant), using the anti-$k_t$~\cite{Cacciari:2008gp} algorithm implementation in \texttt{FastJet}~\cite{Cacciari:2011ma}, with a jet radius $R=1.2$ (different choices for the radius are considered later).  
The clustered jets are required to satisfy $p_{T,J} > 600\text{ GeV}$ and $\left|\eta_J\right| \leq 2.4$.\footnote{\label{foot:bg}Our background processes involving a QCD jet are generated with a parton level cut $p_{T,j} > 500$ GeV to have good statistics in relevant phase space.} After this, the jet with the highest $p_T$ is considered as the candidate diboson jet. Three subjets of this jet are obtained by the N-subjettiness algorithm~\cite{Thaler:2010tr}, using the \texttt{NsubjettinessPlugin} module in \texttt{FastJet}, with the same jet radius and $\beta = 1$ with the axis choice of \texttt{KT Axes}. 
For each of the three subjets, $\text{LSF}$ is calculated for each lepton constituent, and the maximum value over the subjet constituents is taken as the $\text{LSF}$ of the subjet. The subjet with the highest $\text{LSF}$ is identified as the lepton subjet, and its $\text{LSF}$ is the $\LSF{3}$ of the candidate diboson jet. The other two subjets are identified as coming from the hadronic decay of $W$, and are required to have a (groomed) invariant mass in the window $(50, 100)$ GeV and (ungroomed) $\tau_{21} < 0.75$. Grooming is performed by Pruning~\cite{Ellis:2009su} with Cambridge-Aachen algorithm, with $z_{\text{cut}} = 0.1$ and $R_{\text{cut}} = 0.5$. In addition to all this, the candidate diboson jet is required to satisfy (groomed) $m_J > 60$ GeV. 

\begin{figure}[ht]
\centering
\includegraphics[width=0.65\textwidth]{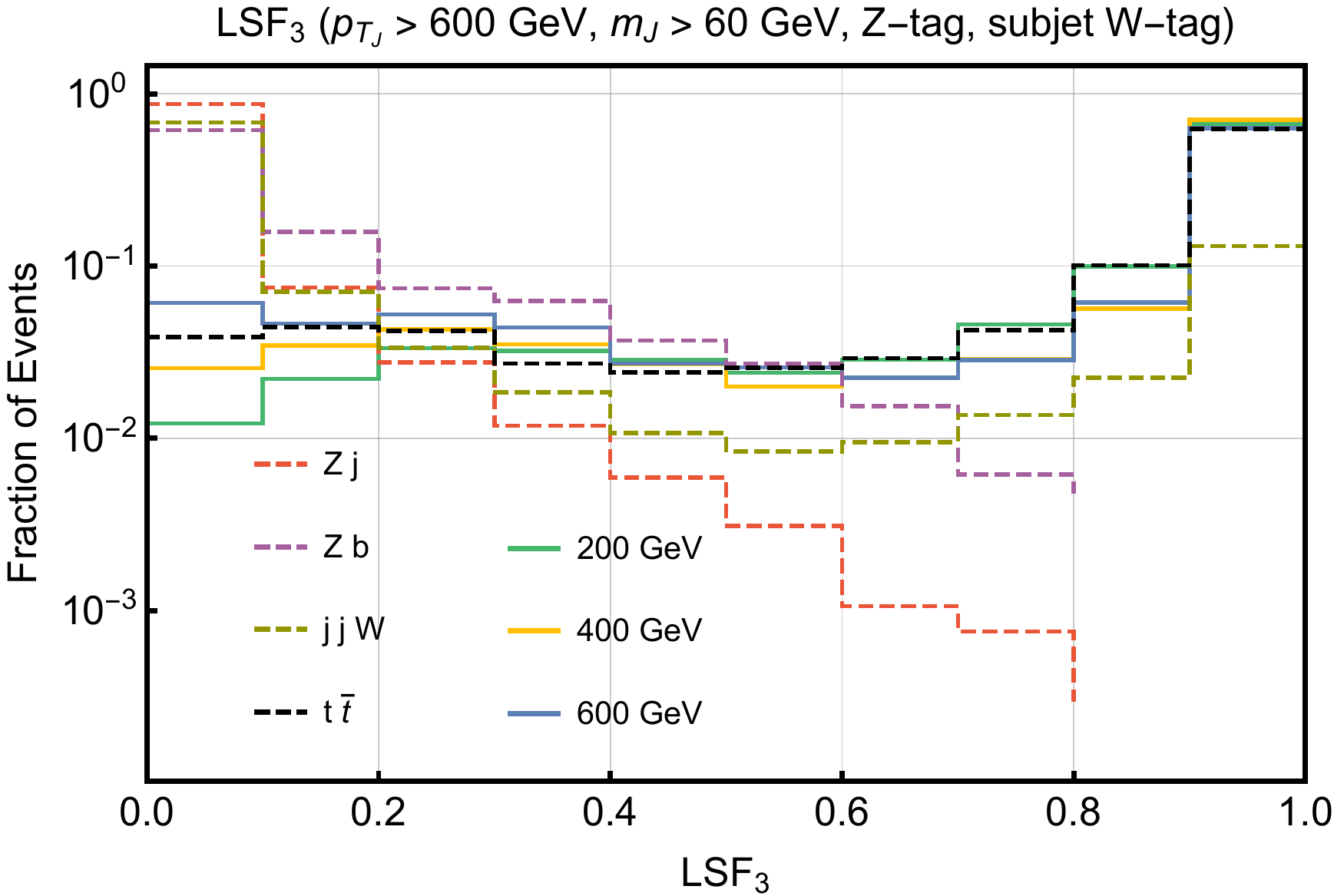}
\caption{
\small{$\text{LSF}_3$ distribution for signal (for a few values of scalar mass) and four backgrounds: $Zj, Zb, jjW$ and $t\bar{t}$.}
}
\label{fig:LSF-Hist}
\end{figure}

We show the distribution of $\LSF{3}$ for the signal and backgrounds in Fig.~\ref{fig:LSF-Hist}, for a few values of the scalar mass. We see that a $\LSF{3}$ cut will be effective at removing backgrounds coming from light jets and $b$-jets while keeping most of the signal, however top-jets and jets with a collinear leptonic $W$ will survive. These could be further reduced by a reconstructed radion mass cut in final states that do not involve any additional sources of missing energy. We will apply a cut of $\LSF{3} > 0.75$ for the semileptonic diboson tagger.

In order to study the dependence of the tagger performance (using a cut on $\LSF{3}$) on the mass of the boosted diboson and on the jet radius, we perform a further study using a scan in the mass range $200 \; \text{GeV} < m_\varphi < 1000 \; \text{GeV}$. The tagger is expected to fail when $m_\varphi$ is sufficiently large that the lepton will frequently fall outside of the jet radius. We therefore compare the performance of the boosted tagger with that of a resolved benchmark strategy which searches separately for a boosted hadronic $W$ boson and a lepton. Most of the steps in this resolved strategy are the same as before. We again require the signal events to have an isolated lepton anti-lepton pair $(\ell^+_z, \ell^-_z)$ of same flavor with an invariant mass in the window $(70, 110)$ GeV which is removed from the analysis. Then, at least one extra isolated lepton is required, which comes from the now well isolated leptonically decaying $W$. Jets are constructed from the track and tower hits with no overlap with $\ell^\pm_z$ (where relevant), again using anti-$k_t$ algorithm implementation in \texttt{FastJet}, with a jet radius $R=1.2$ (different choices for the radius are considered later). The clustered jets are required to satisfy $p_{T,J} > 600\text{ GeV}$, $-2.4 \leq \eta_J \leq 2.4$, and no isolated ($W$) lepton in the jet radius. After this, the highest $p_T$ jet is considered as the candidate diboson jet, and is required to satisfy the standard $W$-jet criteria on (pruned) mass and 2-prongedness variable: $65 \leq m_J \leq 105$ GeV and $\tau_{21} \leq 0.75$. 

Before proceeding, a comment about jet radius is in order. It is important to choose an appropriate jet radius $R$ that should be large enough to capture the decay products from $W$ but small enough to not allow contamination from outside activity. We choose three representative values for $R$: $0.5, 0.8$ and $1.2$, which represent a range of jet radii used by ATLAS/CMS collaborations.

\begin{figure}[ht]
\centering
\begin{subfigure}[b]{0.45\textwidth}
\includegraphics[width=\textwidth]{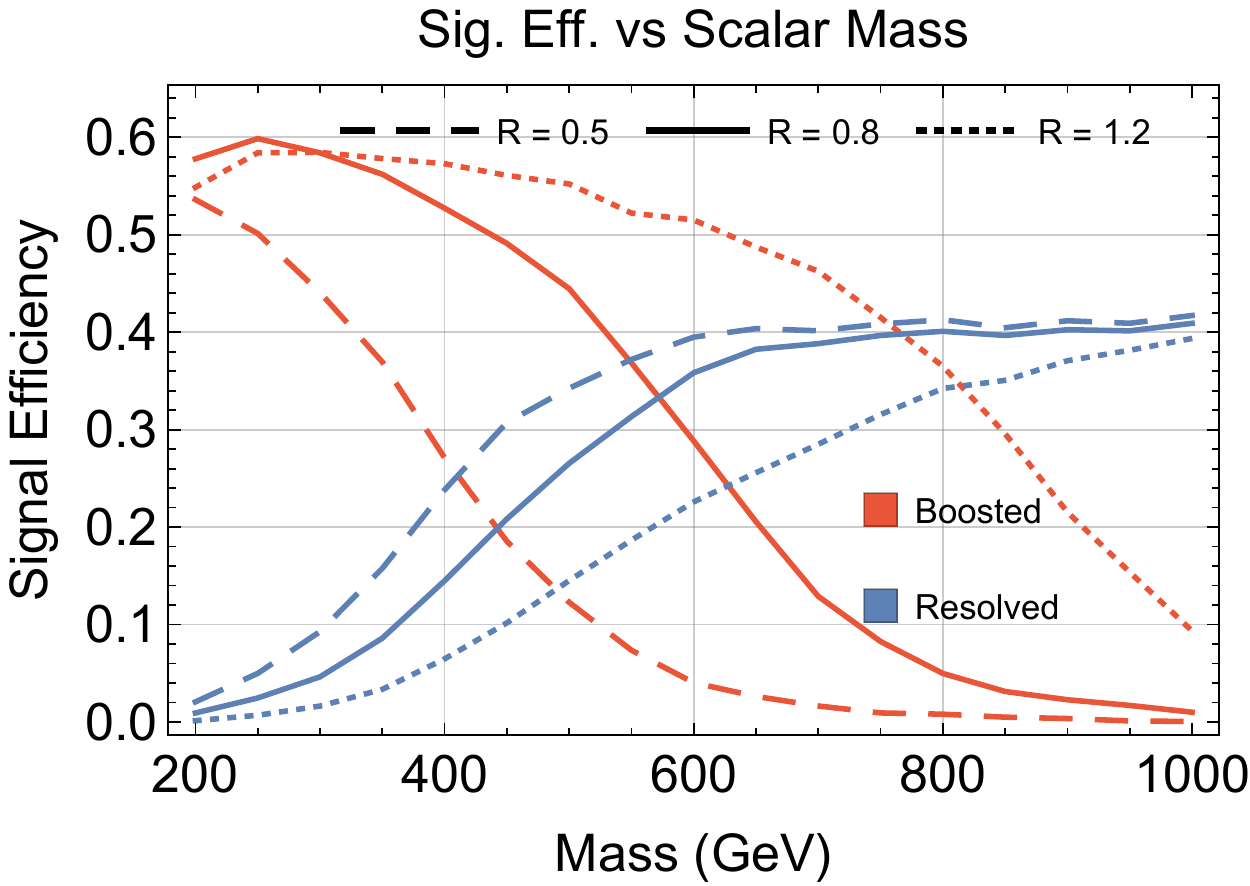}
\caption{}
\label{fig:LSF-Eff-SigEff}
\end{subfigure}
\begin{subfigure}[b]{0.45\textwidth}
\includegraphics[width=\textwidth]{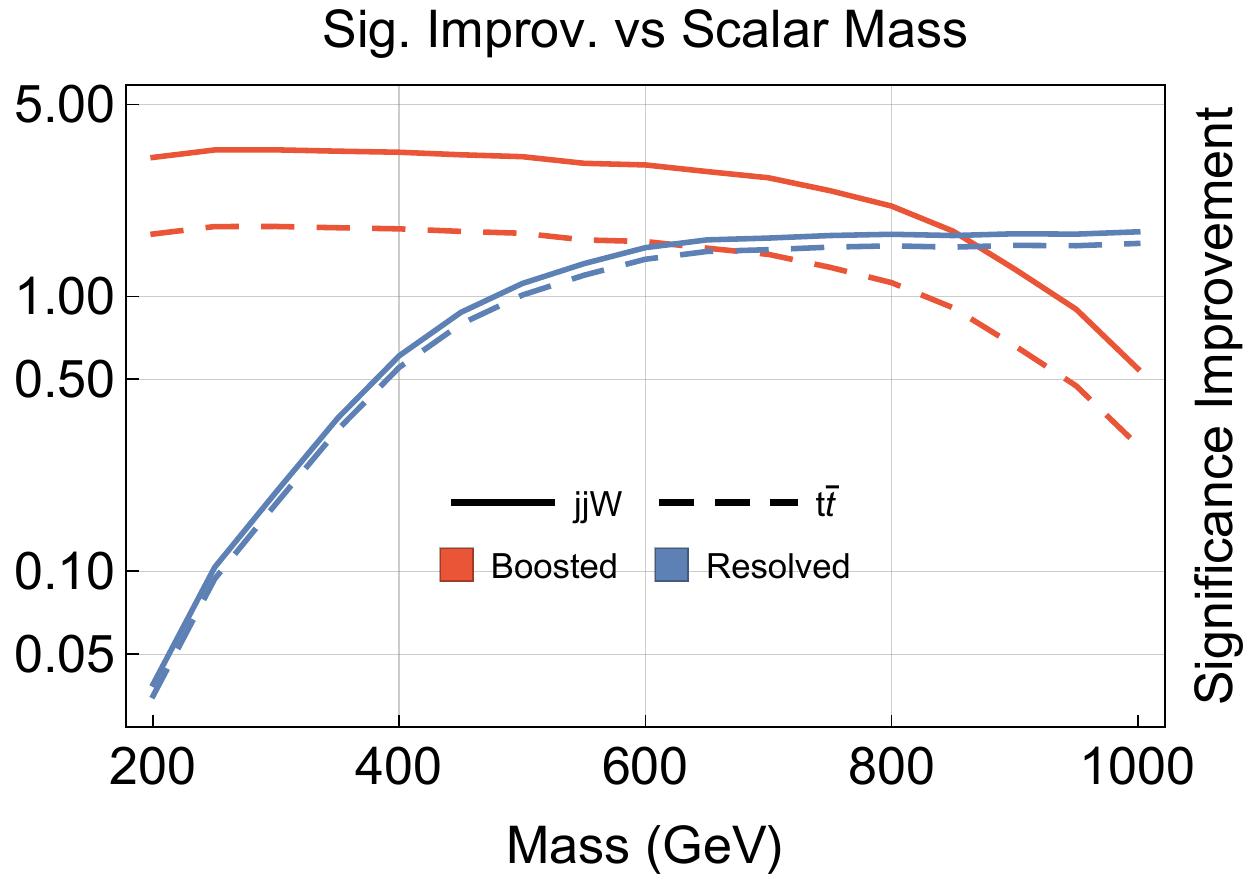}
\caption{}
\label{fig:LSF-Eff-SigImprov}
\end{subfigure}
\caption{
\small{A comparison of the performance of the boosted tagger vs resolved tagger. Left: Signal efficiency as a function of scalar mass, for the boosted tagger, as compared to the resolved tagger. Three values for jet radius are considered, as indicated. Right: Significance Improvement (for the two dominant backgrounds $jjW$ and $t\bar{t}\,$) as a function of scalar mass, for the boosted tagger, as compared to the resolved tagger. The radius for the boosted (resolved) tagger is taken to be 1.2 (0.8). In both cases, the transition happens around 600 GeV.}
}
\label{fig:LSF-Eff}
\end{figure}

We are now in a position to quantify the performance of our semileptonic diboson tagger as a function of scalar mass. Fig.~\ref{fig:LSF-Eff} shows the signal efficiency and significance improvement (for two dominant backgrounds), as a function of the scalar mass. For comparison we also show the performance of the resolved tagger. We show the signal efficiency for three values of the jet radius $R = 0.5, 0.8 \text{ and } 1.2$. For the significance improvement, we take $R=1.2$ for the boosted tagger and $R=0.8$ for the resolved tagger. We see that for low masses, the boosted tagger has $\sim50\%$ efficiency, and it provides a significance improvement of a factor of few for the dominant backgrounds. The performance of both the taggers is counted with respect to the number of events that have a leptonic $Z$ identified (where relevant), and a diboson-jet with $p_{T,J} > 600$ GeV. We notice that both the efficiency and the significance improvement of boosted tagger start dropping around $\sim 600$ GeV, where the resolved tagger starts getting more efficient. As elaborated earlier, at these mass values, the leptonic $W$ from the scalar decay starts getting well separated, so that a boosted strategy becomes less useful.

\section{Application to Warped Framework}
\label{sec:application_to_warped_framework}

Having discussed the tagging algorithms for boosted diboson objects, we are now in a position to discuss specific applications. To this end, we employ the tensor model arising in the context of the extended warped extra-dimensional framework. We first discuss relevant interactions in the model in the next subsection, focusing on the parameter choice for our collider analysis. Our main results will follow in the next subsections, in terms of fully hadronic and semi-leptonic decays of the scalar in this model. 
 
\subsection{Model and parameter choice}
\label{warped_review}
We begin our discussion with a brief description of a model within the warped extra-dimensional framework (see~\cite{Agashe:2016rle, Agashe:2017wss} for more details). The extended warped extra-dimensional setup, which was originally proposed in Ref.~\cite{Agashe:2016rle}, is a natural generalization of the standard Randall-Sundrum model. It is based on the observation that different fields (gravity, gauge, and matter) can propagate different amount into the IR of the bulk, but with non-trivial orderings. Instead of one single bulk, we can then introduce multiple bulks separated by intermediate branes, and this leads to new constructions that allow for more diverse theoretical and phenomenological explorations. For instance, assuming a little hierarchy exists, one can consider a model with two separate bulks (i.e. three branes) with the SM Higgs and fermions constrained to propagate only in the first part of the bulk, between UV and the intermediate brane, while gravity and gauge fields allowed to propagate the entire bulk (from UV to IR brane). Although the model suffers from little hierarchy problem, it can address several things at once -- consistency with flavor/CP violation constraints, absence of signals from the LHC so far, and constraints from EW precision measurements. At the same time it make a very concrete prediction about the form of new physics at the TeV scale, as 5D dual of vectorlike confinement~\cite{Agashe:2016rle}.

\begin{figure}[t]
\centering
\includegraphics[width=0.6\linewidth]{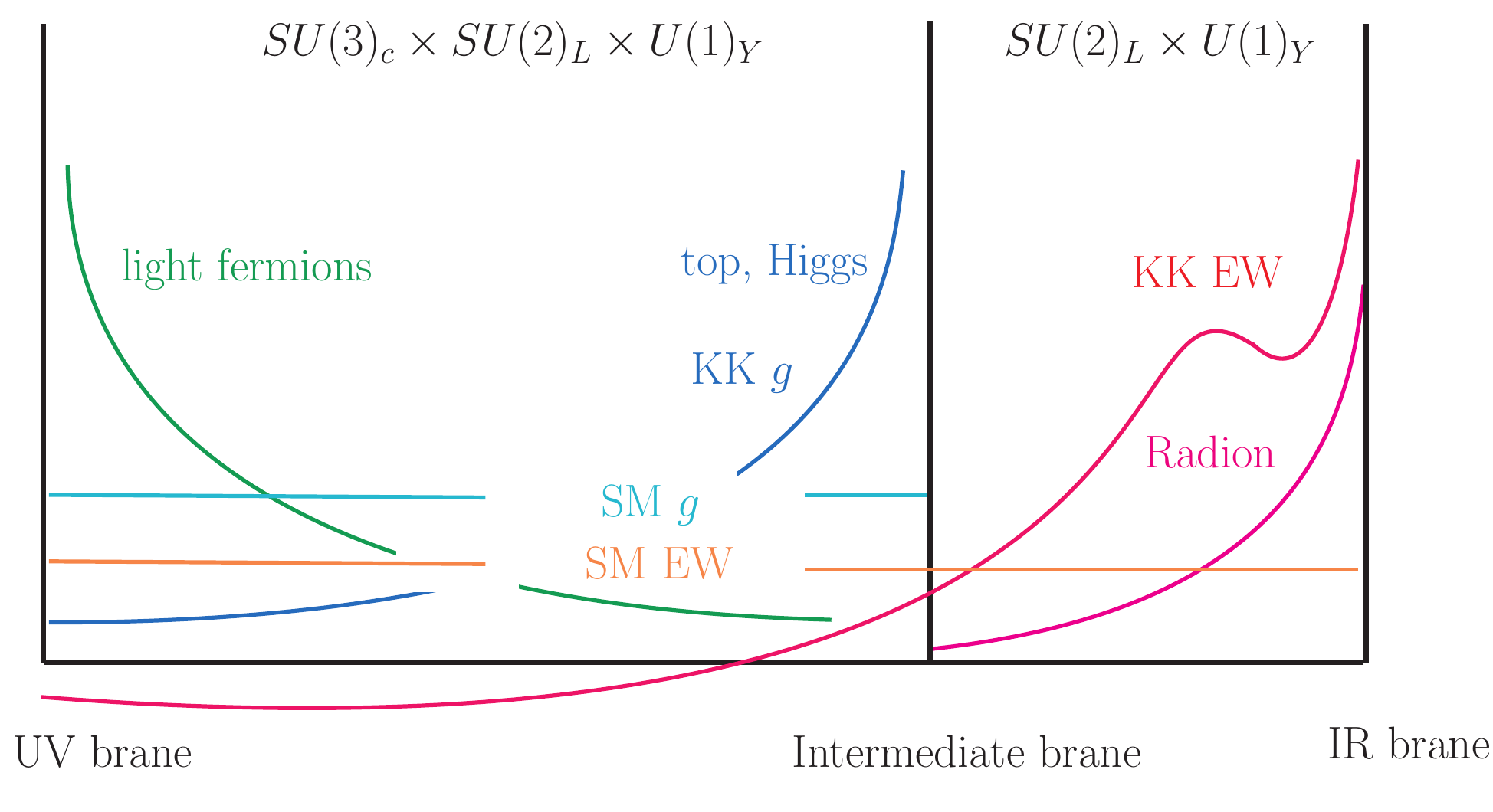}
\caption{\small{
The extended warped extra dimensional model with only SM EW gauge fields in the full bulk (between UV and IR brane), but matter, Higgs and other gauge fields in subspace (between UV and intermediate brane). Schematic shapes of profiles for various particles (zero mode SM fermions and gauge bosons, an (IR) radion, and a generic KK mode) are shown.}
}
\label{fig:3braneTensorModel}
\end{figure}

The tensor model that we study here was first considered in Ref.~\cite{Agashe:2017wss} (see Fig.~\ref{fig:3braneTensorModel}). It again consists of two parts of bulks and only the EW gauge fields and gravity live in the entire bulk, while all the other fields including gluon propagate only down to the intermediate brane. The scale of the intermediate brane is set at $\mathcal{O} (10)$ TeV to be consistent with flavor/CP constraints, while the IR brane scale is taken as $\mathcal{O} (1)$ TeV. The new particles relevant for LHC phenomenology are KK EW gauge bosons and a radion $\varphi$ that describes the fluctuation of the IR brane.\footnote{The radion associated with the fluctuation of the intermediate brane is assumed to be heavy.} The interactions of these new particles with SM particles and/or among themselves can be described by the following Lagrangian:
\begin{align}
\mathcal L^{\rm EW}_{\rm warped} 
&\ni
-\frac{1}{4} \left( V_{\mu\nu} \right)^2 - \frac{1}{4} \left( V^{\rm KK}_{\mu\nu} \right)^2 + \frac{1}{2} m_{\rm KK}^2 \left( V^{\rm KK}_\mu \right)^2 \nonumber \\
&+
\frac{g^2_V}{g_{V_{\rm KK}}} V_{\rm KK} ^\mu J_{V\mu} + \left(  -\frac{1}{4}\frac{g_{\rm grav}}{g_{V_{\rm KK}}^2} g_{V}^2  V_{\mu\nu}V^{\mu\nu}+ \epsilon \frac{g_{\rm grav}}{g_{V_{\rm KK}}^2} g_{V_{\rm KK}} g_{V} V_{\mu\nu}V^{\mu\nu}_{\rm KK} \right) \frac{\varphi}{m_{\rm KK}}\,, \label{eq:warped_model_couplings}
\end{align}
where we used the notation $V_{\rm KK}\in \{W_{\rm KK}^{1,2,3},\,B_{\rm KK}\}$ to denote KK EW gauge bosons collectively. Similarly for SM EW gauge bosons: $V \in \{W^{1,2,3},\,B\}$. Corresponding gauge couplings are denoted as $g_{V_{\rm KK}}$ and $g_{V}$, respectively. We assume all KK EW gauge bosons have the same mass $m_{\rm KK}$. Here $g_{\rm grav}$ is the coupling associated with KK graviton and radion.

Now we discuss each interaction term one by one. The first term in the second line of Eq.~(\ref{eq:warped_model_couplings}) describes the coupling between KK EW gauge bosons and SM matter fields (fermions and Higgs). Specifically, KK EW gauge boson couples to SM matter fields via its coupling to SM matter current $J_{V\mu}$ associated with the corresponding SM EW gauge boson $V$. This coupling enables production of KK EW gauge bosons through $q\bar{q}$ annihilation and also they can decay to a pair of SM matter fields, e.g. dilepton or dijet. The second term in the second line is what we discussed in section~\ref{bound}: coupling of a scalar $\varphi$ to a pair of SM EW gauge bosons. In fact, we see that the current warped model we study here generates the couplings with Wilson coefficients of $A_W = \frac{1}{4}\frac{g_{\rm grav}}{g_{W_{\rm KK}}^2}$ and $A_B = \frac{1}{4}\frac{g_{\rm grav}}{g_{B_{\rm KK}}^2}$. Also, the UV scale $\Lambda$ in Eq.~(\ref{tensor1}) is identified with $m_{\rm KK}$. The last interaction, third term in the second line, is the coupling among $V_{\rm KK} - V - \varphi$. It is a consequence of radius stabilization and this may be seen from the appearance of the parameter $\epsilon$, which, in dual CFT picture, corresponds to the anomalous dimension of the scalar operator (Goldberger-Wise operator) needed for the stabilization. This coupling allows for KK gauge boson to decay to a SM gauge boson and a radion. In this study, for concreteness, we fix the couplings in Eq.~(\ref{eq:warped_model_couplings}) to be the same as those in Ref.~\cite{Agashe:2017wss}: $g_{W_{\rm KK}}=3$, $g_{B_{\rm KK}}=6$,
$\epsilon=0.5$ and $g_{\rm grav}=6$.
 
The signal process of interest begins with single production of KK $W$ boson via $q\bar{q}'$ annihilation dictated by the first coupling in Eq.~\eqref{eq:warped_model_couplings}.
The $W_{\rm KK}$ subsequently decays to a SM $W$ gauge boson and a radion $\varphi$ via the last coupling in Eq.~\eqref{eq:warped_model_couplings}, followed by a further decay of $\varphi$ to a pair of SM $W$ gauge bosons via the second coupling in Eq.~\eqref{eq:warped_model_couplings}. 
We therefore have three $W$ gauge bosons in the final state --- a ``prompt'' $W$ (from the direct decay of $W_{\rm KK}$) and two ``secondary'' $W$'s (from the $\varphi$ decay), as shown in Fig.~\ref{fig:scenario}.
The mass ranges we consider in our collider study are $[2000,4000]$ GeV for $W_{\rm KK}$ and $[200,900]$ GeV for $\varphi$.
In much of the parameter space the radion from the KK $W$ decay will be boosted, and in turn, the two $W$'s from the radion decay will be merged to form a boosted diboson. In fact, the production rate for the radion via the decay of a KK gauge boson is much greater than the direct production rate via VBF or associated production via an off-shell SM gauge boson. For example, we find the ratio $\sigma_\text{boosted} / \sigma_\text{direct} \simeq 6$ for $m_{W_\text{KK}} = 3 \; \text{TeV}, m_\varphi = 500 \; \text{GeV}$, and generally $\sigma_\text{boosted} / \sigma_\text{direct} \sim O(10)$ over the considered parameter space. This means that searches dedicated to the boosted production of the radion are more likely to provide a discovery of the particle than those dedicated to its production just above threshold in VBF and associated production.

The constraints on the model coming from existing LHC searches can be separated into three categories. First there are the dedicated searches for multi-TeV resonances decaying into SM particles, which will be sensitive to the $W_\text{KK}$ and other KK gauge bosons. Second are the searches for electroweak-scale resonances, and third are studies of multi-boson processes, both of which were discussed in section~\ref{bound} and which will be revisited in this section.

Of the multi-TeV resonance searches, the most sensitive to this model are the searches for $W' \to \ell \nu$, which search for a transverse mass peak using the momentum of a high-$p_T$ lepton and $E_{T}^{\rm miss}$. Using the search of Ref.~\cite{ATLAS-CONF-2018-017} which uses $80 \; \text{fb}^{-1}$ of 13~TeV data we find that this constrains $m_{\rm KK} \gtrsim 3 \; \text{TeV}$, as indicated more precisely by the purple lines in Figs.~\ref{fig:FullyHadronic-ExclusionAndDiscoveryReach} and \ref{fig:SemiLeptonic-ExclusionAndDiscoveryReach}. Dijet resonance searches such as Refs.~\cite{Sirunyan:2018xlo, Aaboud:2017yvp} are sensitive to the decay $W_\text{KK} \to q\bar{q}'$. But while this branching ratio is three times larger than in to dileptons (due to the color factor), the limits on the leptonic channel are two orders of magnitude more powerful because of the cleaner final state. Searches for resonances decaying into $t \bar{b}$ such as Refs.~\cite{Sirunyan:2017vkm, Aaboud:2018juj} have similar sensitivity as the dijet searches. Finally, Ref.~\cite{Sirunyan:2017acf} contains a search for excited quark resonances decaying into $qW$ with a hadronically decaying $W$ boson. In this search, the ``quark jet'' $q$ is defined as a radius $R = 0.8$ jet which fails the substructure and jet-mass identification criteria for a $W$ or $Z$ boson. The signal process $W_\text{KK} \to \varphi W$, with $\varphi \to W W \to qqqq$ will be visible as a resonance in this search with the $\varphi$ jet being identified as the ``quark jet'' in the case that the decay products of the $\varphi$ are contained in an $R = 0.8$ jet, which will have high efficiency in the limit that $m_\varphi / m_{W_\text{KK}} \ll 0.2$. For heavier radion masses much of the energy will not be captured in the jet radius. This will result in a smearing of the resonance peak which might absorbed in the background fit of the search~\cite{Aguilar-Saavedra:2017iso}. A detailed analysis of this behavior is beyond the scope of this paper; nonetheless, we can conservatively estimate the maximum possible sensitivity of this search by assuming 100\% reconstruction efficiency for the $\varphi W$ resonance as a $qW$ resonance. We find that the limits are at best 80(2) fb for $m_{W_\text{KK}} = 2(4) \; \text{TeV}$, while our model predicts theses cross-sections to be 20(0.2) fb. Therefore this search is significantly weaker than $W' \to \ell \nu$.

Searches targeting multi-TeV diboson resonances~\cite{Aaboud:2017eta, Aaboud:2017fgj, Aaboud:2017itg,Sirunyan:2017acf, Sirunyan:2018iff, Sirunyan:2018ivv,Sirunyan:2018hsl} may have some sensitivity to the cascade decays of $V_\text{KK} \to V \varphi \to V(VV)$. As for the $qW$ resonance search above the signature will depend sensitively on the overlap of the vector bosons, and the combinatoric ambiguity in vector-boson selection will typically give rise to broad multi-TeV features in the case that the vector bosons are sufficiently well separated to be tagged. We refer to Refs.~\cite{Aguilar-Saavedra:2017iso, Aguilar-Saavedra:2016xuc} for further discussion of these issues. The fact that the boosted diboson signal is quite different than those targeted by the diboson searches motivates the use of boosted diboson taggers in a dedicated search, as will be described later in this section.

Now we come to searches dedicated to electroweak-scale resonances. Since we anticipate that the discovery of the KK gauge bosons would be made first in the $W' \to \ell \nu$ search, we focus here on searches that could provide clear evidence of the radion resonance.

\begin{figure}[t]
\centering
\includegraphics[width=0.6\linewidth]{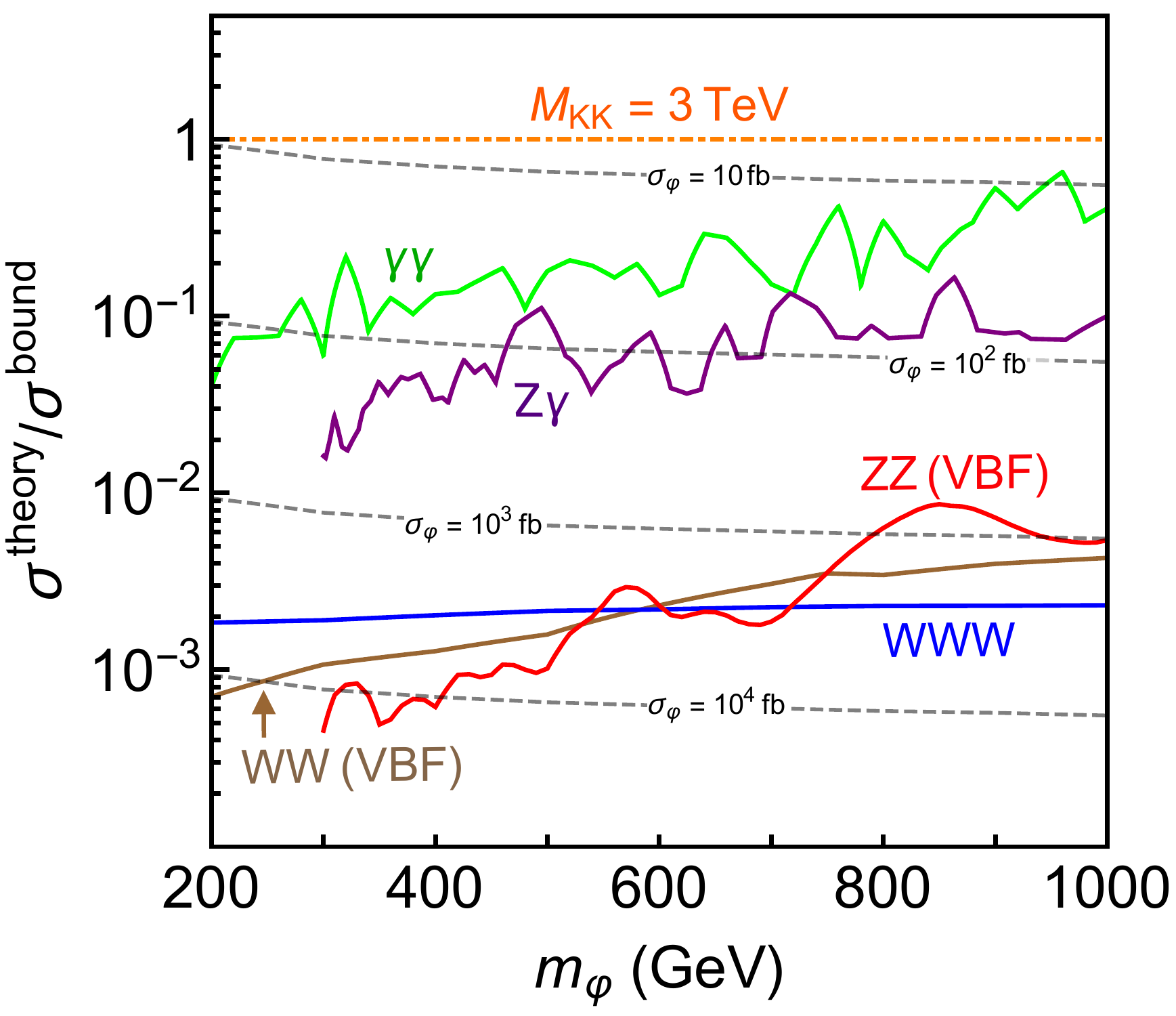}
\caption{\small{
The ratio of cross section predicted by the warped model to the experimental bounds as a function of $m_\varphi$. Five relevant searches are shown: ATLAS diphoton (green), ATLAS diboson $WW$ (fully leptonic, in brown), ATLAS diboson $ZZ$ ($\ell \ell q q  + \nu \nu q q$ in red), CMS $Z \gamma$ (leptonic, in purple), and ATLAS triboson $WWW$ (blue). Regions below the solid lines are excluded by the corresponding experiments. The dashed lines show the scalar/radion production cross section $\sigma(p p \to \varphi)$, sum of both VBF and associated production. For $ZZ$ (red) and $Z\gamma$ (purple), results are shown only down to 300 GeV because relevant experimental search results are reported only to that point. The orange dashed dot line indicates our benchmark point with $m_{\rm KK} = 3 \; \text{TeV}$.}
}
\label{fig:bound_warped_model}
\end{figure}

We found in section~\ref{bound} that the strongest bounds on direct production of the scalar in the tensor model came from diphoton resonance searches due to the significant branching fraction of $\varphi$ to photons in this class of model. In the warped model, the processes $V_\text{KK} \to V \varphi \to V (\gamma \gamma)$ will also result in a clean diphoton peak at $m_\varphi$ due to the inclusive nature of these searches, and will in fact provide the dominant contribution due to their large cross section. Because the signal topology is significantly different than that assumed in the diphoton searches, recasting the limits requires estimating the selection efficiency for the boosted production. We do this with a parton-level simulation using \texttt{MadGraph5@aMC}, applying the cuts and identification efficiencies described in Ref.~\cite{Aaboud:2017yyg}. We find that for $m_{W_\text{KK}} = 3 \; \text{TeV}$ this selection efficiency ranges between 0.4 and 0.7. The photon isolation cut of $\Delta R_{\gamma\gamma} > 0.4$ is relevant only for $m_\varphi \approx 200 \text{ GeV}$ where it has an efficiency around 0.7. The resulting exclusion sensitivity is given by the green line in Fig.~\ref{fig:bound_warped_model}. It is worth noting that since the majority of these diphoton events come in a boosted configuration, some relatively minor modifications to the search could substantially reduce the background for the diphoton resonance. This could include a cut on the $p_T$ of the diphoton system, and using a narrower photon isolation cone which would improve signal efficiency at low radion masses.

Similarly, the processes $V_\text{KK} \to V \varphi \to V (Z \gamma)$ will give a clean peak at $m_\varphi$ in the $\ell \ell \gamma$ channel of Ref.~\cite{Sirunyan:2017hsb}, so long as the prompt $V$ does not decay into charged leptons. Again, due to the substantially different signal topology, we calculate the efficiency for our model using parton level simulations to implement the cuts used in the analysis in order to obtain the bounds given by the purple line in Fig.~\ref{fig:bound_warped_model}. Due to the relatively small signal fraction of $Z(Z \gamma)$ events, there may also be a small high-mass feature due to the pairing of a leptonically decaying prompt $Z$ with the photon in the case that the $Z$ from the radion decays invisibly, or if it decays hadronically and the resulting jets do not overlap with the photon. This case requires a dedicated study, which is beyond the scope of our present work, but these complications also motivate a dedicated search for this kind of scenario in order to identify the radion.

The relatively large branching fractions into $\gamma \gamma$ and $Z \gamma$ is characteristic of the model with tensor-type couplings, but would not be expected in the case that the scalar were to decay via a Higgs portal type coupling. It is therefore worth also considering the diboson and triboson searches which are more directly connected with the minimally assumed couplings of the diboson resonance to $W$ and $Z$. The searches for diboson resonances produced via VBF which were discussed in section~\ref{bound} are also relevant here, and we present the sensitivity of these in the brown and red lines of Fig.~\ref{fig:bound_warped_model}. Finally, the third category of constraint comes from the $WWW$ search of Ref.~\cite{Aaboud:2016ftt}. We recast this search as described in section~\ref{bound}, resulting in the constraint given by the blue line in Fig.~\ref{fig:bound_warped_model}.

\subsection{Results}
\label{warped_results}
Finally, in this section we investigate the performance of dedicated search strategies using the taggers discussed in section~\ref{strategy} for the model described above. The details of the search will vary according to the combination of leptonic and hadronic decays of the three $W$ boson in the event. In the cases where the radion decays fully hadronically, $\varphi \to WW \to 4q$, the taggers of section \ref{hadronic_tagger} can be used to fully reconstruct the radion. The $W_\text{KK}$ can then be reconstructed using the radion and the prompt $W$ whether it decays hadronically or leptonically. In the case that the radion decays semi-leptonically, $\varphi \to WW \to \ell \nu qq$ it can be reconstructed using the tagger of section \ref{semileptonic_tagger}. The $W_\text{KK}$ can then be reconstructed only in the case that the prompt $W$ decays hadronically. We therefore consider the final states $qq(qqqq),~\ell \nu (qqqq),~qq(\ell \nu qq)$ with the objects in the parentheses denoting the decay products of $\varphi$. These have branching fractions of 0.3, 0.1, and 0.2 respectively.

In all cases, the process $W_\text{KK} \to W \varphi$ gives rise to two back-to-back objects which will reconstruct the $W_\text{KK}$ resonance. The general strategy will therefore be to tag a boosted $W$ recoiling from a boosted $\varphi$, and search for a bump in the invariant masses of the reconstructed $\varphi$ and $W_\text{KK}$ on top of the Standard Model backgrounds which will form a smooth distribution in this plane. We implement a simplified version of such a bump-hunt by selecting events within windows of $m_\varphi$ and $m_{W_\text{KK}}$, and comparing the number of simulated signal and background events. We scan these windows in the $m_\varphi - m_{W_\text{KK}}$ plane in order to determine a projected exclusion and discovery reach at $300 \text{ fb}^{-1}$.

For exclusion, we compute the Poisson Likelihood ratio $\lambda_\text{exc} = \mathcal{L}\left(s+b | b\right) / \mathcal{L}\left(b | b\right)$ and exclude any signal which results in $\sqrt{- 2 \log \lambda_\text{exc}} > 2$ (using the asymptotic formulae of~\cite{Cowan:2010js}). For projected discovery, we compute the Poisson Likelihood ratio $\lambda_\text{dis} = \mathcal{L}\left(b | s+b\right) / \mathcal{L}\left(s+b | s+b\right)$ and consider within discovery reach any signal which results in $\sqrt{- 2 \log \lambda_\text{dis}} > 5$. These estimates ignore any potential systematic uncertainties, and so our projections must be regarded only as rough estimates.

\subsubsection{Fully hadronic radion decay}
\label{sec:warped_results_fullyhad}

For the channels involving a fully hadronic radion decay, $qq(qqqq)$ and $\ell \nu (qqqq)$, we simulate $10,000$ signal events for each channel and for each mass point on a grid in the plane $2 \; \text{TeV} < m_{W_\text{KK}} < 4 \; \text{TeV}$ and $200 \; \text{GeV} < m_\varphi < 900 \; \text{GeV}$, again using \texttt{MadGraph5@aMC}. 
The dominant background to the fully hadronic channel is QCD dijets, while for the $\ell \nu (qqqq)$ channel the main background is $W$ + jets. In the following we describe the detailed analysis steps taken for each channel separately, and then illustrate the reach when this analysis routine is used to perform a mass scan in the $(m_{W_\text{KK}}, m_\varphi)$ plane.

For the QCD dijet background, we simulated 20 million events in $2j$ and $3j$ samples which were matched using the shower-kt scheme~\cite{Alwall:2008qv}. In order to generate sufficient statistics on the high energy tail, these events were simulated with a biased weight of $(m_\text{eff} / m_0)^6$, where $m_\text{eff}$ is the invariant mass of all partonic objects and $m_0$ is an arbitrary reference mass scale. We also require cuts of $H_T > 800 \; \text{GeV}$, $p_{T, j_1} > 200 \; \text{GeV}$, and $|\eta_j| < 4$ at generation level. Background and signal events are showered with \texttt{Pythia8} and detector response is simulated with \texttt{Delphes3}~\cite{deFavereau:2013fsa}. 
The particle flow output of \texttt{Delphes3} is used to perform our analysis in the $qq(qqqq)$ channel. We cluster jets using the anti-$k_t$ algorithm with radius parameter $R = 1.2$, and require these jets to satisfy $|\eta| < 2$. We select the hardest two such jets as our radion candidate and $W$ candidate. Each of these jets are required to have $p_T > 400 \; \text{GeV}$, and the jet pair must have $p_{T, W} + p_{T, \, \varphi} > 1 \; \text{TeV}$. The more massive jet is taken to be the radion candidate, and the lighter is the $W$ candidate. The $W$ candidate is groomed using the Soft Drop algorithm, with $\beta = 0$ and $z_\text{cut} = 0.1$. We require that this jet has $\tau_{21} < 0.6$ (calculated using the ungroomed jet) and $65 \; \text{GeV} < m_J < 95 \; \text{GeV}$, consistent with a jet originating from a boosted $W$-boson. For the radion jet, we consider both the low-mass hypothesis and the intermediate-mass hypothesis, in which case we groom and tag the jet with the merged tagger and the intermediate tagger, respectively. 

For the $W$+jets background, we simulate 2 million $Wj$ and $Wjj$ events, which are matched again using the shower-kt scheme. These are simulated with the same biased weight as for the fully hadronic sample above. We impose parton level cuts of $p_T > 100 \; \text{GeV}$ on the hardest jet, $H_T > 250 \; \text{GeV}$, $|\eta_j| < 3.5$, $p_{T, \, \ell} > 40 \; \text{GeV}$, $|\eta_\ell| < 2.5$, and $E_{T}^{\text{miss}} > 10 \; \text{GeV}$. These events and the $\ell \nu (qqqq)$ signal events are also passed through \texttt{Pythia8} and \texttt{Delphes3}, and the particle-flow output is used for analysis. Jets are clustered using anti-$k_t$ with radius parameter $R=1.2$, and are required to have $|\eta| < 2$. Jets that overlap with isolated leptons ($\Delta R_{J \ell} < 1.2$) are removed from the sample. The hardest remaining jet is considered the radion-candidate, and this jet is required to pass either the merged or intermediate radion tagging criteria. The event is required to have $E_{T}^{\text{miss}} > 60 \; \text{GeV}$ and to contain at least one charged lepton (electron or muon) with $p_T > 50 \; \text{GeV}$; the hardest such lepton is assumed to come from the decay of a boosted $W$ boson. The neutrino momentum is reconstructed by solving the quadratic equation
\begin{align}
m_W^2 = (p_{\nu}+p_\ell)^2 = 2\left(E_\ell \sqrt{|\vec{P}_T^{\textrm{miss}}|^2+p_{z,\nu}^2}-\vec{p}_{T,\ell}\cdot \vec{P}_T^{\textrm{miss}} - p_{z,\ell}p_{z,\nu}\right).
\end{align}
In the case that two real solutions are found we take the smaller solution. In the case of complex solutions, we take the neutrino momentum to be given by the real part.\footnote{This is equivalent to assuming the neutrino is parallel to the charged lepton.} This allows us to reconstruct the four-momentum of a $W$-candidate, $p_W = p_\ell + p_\nu$. We require both $p_{T, \, W},  p_{T, \, \varphi} > 400 \; \text{GeV}$, as well as $|\eta_W| < 2$.

\begin{figure}[t]
\centering
\begin{subfigure}[b]{0.45\textwidth}
\includegraphics[width=\textwidth]{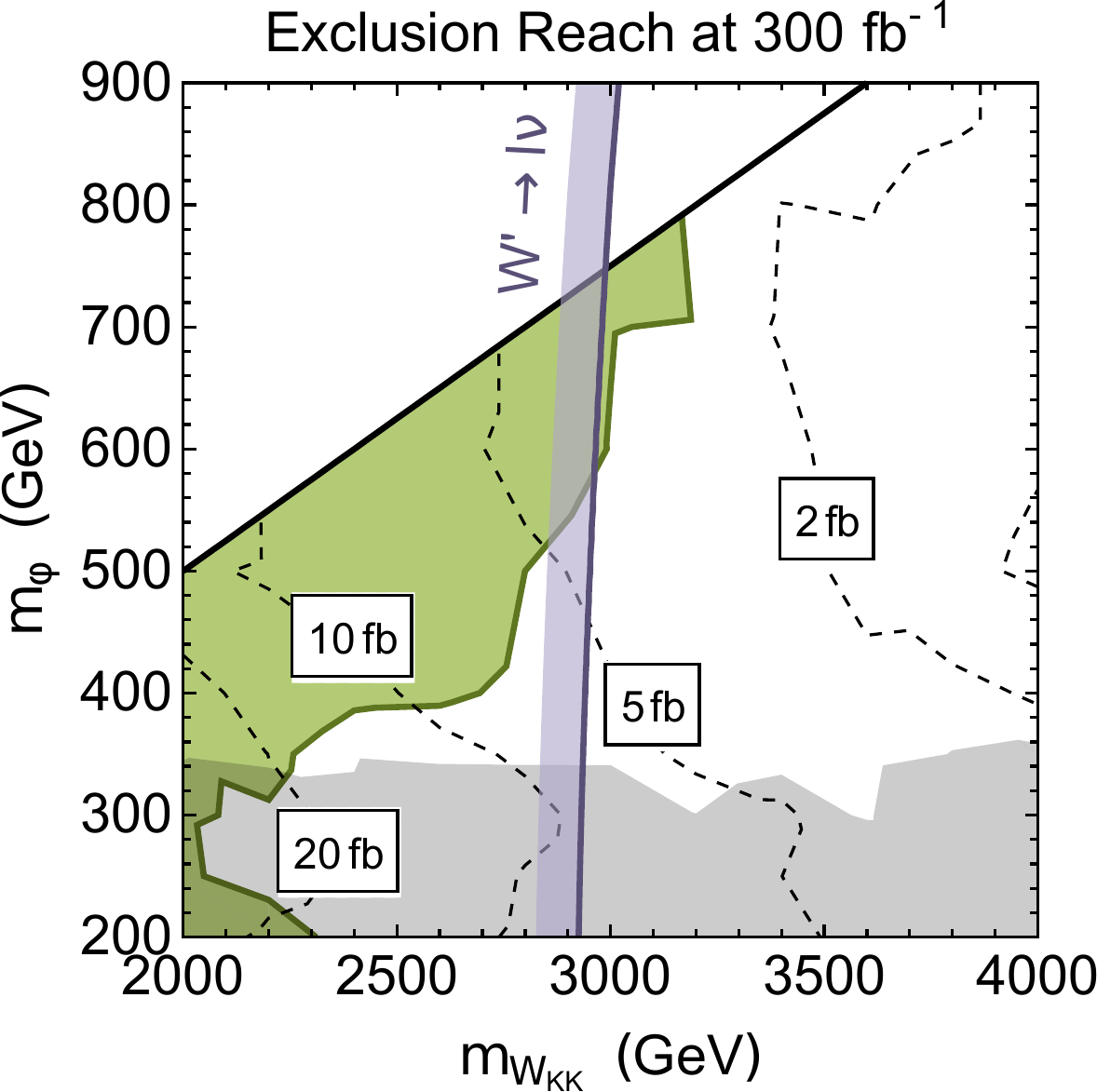}
\caption{\small{Hadronic prompt $W$}}
\label{fig:Hadronic-hadronic}
\end{subfigure}
\begin{subfigure}[b]{0.45\textwidth}
\includegraphics[width=\textwidth]{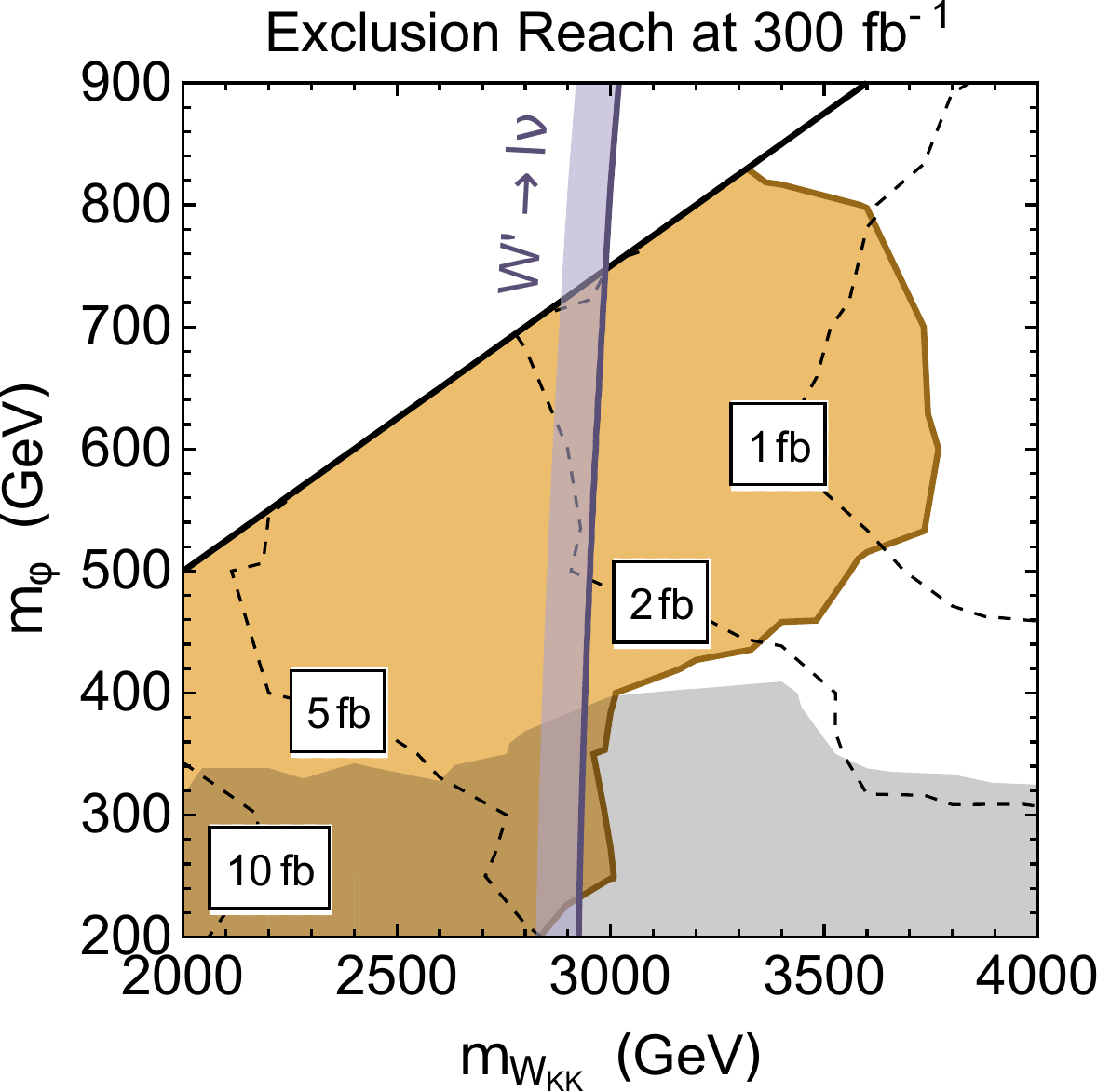}
\caption{\small{Leptonic prompt $W$}}
\label{fig:Hadronic-leptonic}
\end{subfigure}
\caption{
\small{Projected exclusion reach with $\mathcal{L} = 300 \; \text{fb}^{-1}$ for a fully hadronic boosted diboson jet recoiling from a hadronic (green) or leptonic (orange) prompt $W$. The purple line indicates the current exclusion due to the search for $W' \to \ell \nu$ of Ref.~\cite{ATLAS-CONF-2018-017}. The dashed contours indicate the limit on $\sigma(W_\text{KK}) \times  \text{BR}(W_\text{KK} \to W \varphi) \times \text{BR}(\varphi \to WW)$ obtained by the proposed search. In all cases, the tagger (intermediate or fully merged) with strongest sensitivity is used. The gray shaded regions below $m_\varphi \simeq 350 \; \text{GeV}$ indicate the region where the fully merged tagger is the most powerful.}
}
\label{fig:FullyHadronic-ExclusionAndDiscoveryReach}
\end{figure}

In each channel we perform a mass scan in the $(m_{W_\text{KK}}, m_\varphi)$ plane. For each $(m_{W_\text{KK}}, m_\varphi)$ hypothesis, we require that the radion-candidate jet mass falls in the window $0.845 \, m_\varphi < m_{J, \, \varphi} < 1.095 \, m_\varphi$ and that the two-object pair ($JJ$ in the $qq(qqqq)$ channel and $J (\ell \nu)$ in the $\ell \nu (qqqq)$ channel) has invariant mass $0.845 \, m_{W_\text{KK}} < m_{JJ} < 1.095 \, m_{W_\text{KK}}$ (these mass windows were obtained empirically from the distributions). The 2$\sigma$ exclusion reach with $300 \;\text{fb}^{-1}$ is illustrated in Figs.~\ref{fig:Hadronic-hadronic} and~\ref{fig:Hadronic-leptonic} for the $qq(qqqq)$ and $\ell \nu (qqqq)$ channels respectively. We find that the search in the $\ell \nu (qqqq)$ is particularly sensitive, with the potential to exclude $W_\text{KK}$ masses as high as 3.7~TeV. We find that the transition between the merged and the intermediate tagger having best sensitivity (indicated by the shaded gray regions) is $m_\varphi \simeq 350 \; \text{GeV}$ independent of $m_{W_\text{KK}}$, consistent with the expectations from the approximate formulae of section \ref{strategy}.

\subsubsection{Semi-leptonic radion decay}
\label{sec:warped_results_semilep}
For the channel with a semi-leptonic radion decay and a fully hadronic prompt $W$ decay, giving the final state $qq(\ell\nu qq)$, we simulate $50,000$ signal events for each mass point on a 2D mass grid, for $m_{W_{KK}} \in [2000, 4000] \text{ GeV}$ and $m_\varphi \in [200, 900] \text{ GeV}$. The relevant SM backgrounds for this signal are: $pp\rightarrow jj,\:\:pp \rightarrow b\bar{b}, \:\:pp \rightarrow jjW$ and $pp\rightarrow t\bar{t}$. Similar to the approach in section~\ref{strategy}, we distinguish between the light quark jets from $b$-jets, to better understand the effect of various cuts on the background. To focus on the relevant part of the phase space as well as have enough statistics, we generate signal and background events requiring at least one jet with $p_T > 500$ GeV. For $jjW$ and $t\bar{t}$ events, we additionally require at least one charged lepton with $p_T > 50$ GeV. We also set $\Delta R_{jj}^{\text{min}} = 0 $ and $\Delta R_{jl}^{\text{min}} = 0$ to include non-isolated leptons and merged jets in our MC sample. 

Jets are constructed from the track and tower hits, using anti-$k_t$ algorithm implementation in \texttt{FastJet}, with a jet radius $R = 1.2$.  The clustered jets are required to satisfy $\left|\eta\right| \leq 2.4$. 
After this, the jets are tagged as $W$-jet or non $W$-jet, depending on the standard $W$-tagging criteria: (groomed) $m_J \in [65, 105]$ GeV and (ungroomed) $\tau_{21} < 0.75$. 
A $W$-tagged jet with the highest transverse momentum is taken as the candidate jet from the hadronic decay of the prompt $W$, while the hardest $p_T$ non $W$-tagged jet is taken as the candidate jet from the semi-leptonic radion decay. For this candidate radion jet, three constituent subjets are obtained again by \texttt{NsubjettinessPlugin} module in \texttt{FastJet}, with the same jet radius ($R=1.2$) and $\beta = 1$, for the axis choice of \texttt{KT Axes}. 
For each of the three subjets, $\text{LSF}$ is calculated for each lepton constituent, and the maximum value over the subjet constituents is taken as the $\text{LSF}$ of the subjet. The subjet with the highest $\text{LSF}$ is identified as the lepton subjet, and its $\text{LSF}$ is the $\LSF{3}$ of the candidate radion jet, required to satisfy $\LSF{3} > 0.75$. We also require that the lepton has $p_{T, \ell} > 100 \; \text{GeV}$. The other two subjets are identified as coming from the hadronic decay of $W$, and are required to have a (groomed) invariant mass in the window $[65, 105]$ GeV and (ungroomed) $\tau_{21} < 0.75$. Grooming is performed by Pruning~\cite{Ellis:2009su} with Cambridge-Aachen algorithm, with $z_{\text{cut}} = 0.1$ and $R_{\text{cut}} = 0.5$. Finally, we reconstruct the neutrino four-momentum according to the procedure described in the previous section, using the identified lepton subjet (inside the radion jet), consistent with coming from an on-shell $W$ decay. 

After this initial curation of events, we impose further cuts to isolate the signal. We require that $p_{T, W} > 700 \; \text{GeV}$, $p_{T, J_\varphi} > 300 \; \text{GeV}$ (where  $J_\varphi$ is the ``radion jet'' without using the reconstructed neutrino), and $E_{T}^{\rm miss} > 100 \; \text{GeV}$. The mass cut on the hadronic $W$-jet inside the radion is taken as $m_J \in [65, 100]$ GeV. Using the reconstructed neutrino four-momentum, we reconstruct the radion and the $W_{\rm KK}$ four-momenta. Next, a window mass cut is imposed on radion and $W_{\text{KK}}$, centered on their mass and with a half width of $20\%$ of their mass.

\begin{figure}[t]
\centering
\begin{subfigure}[b]{0.45\textwidth}
\includegraphics[width=\textwidth]{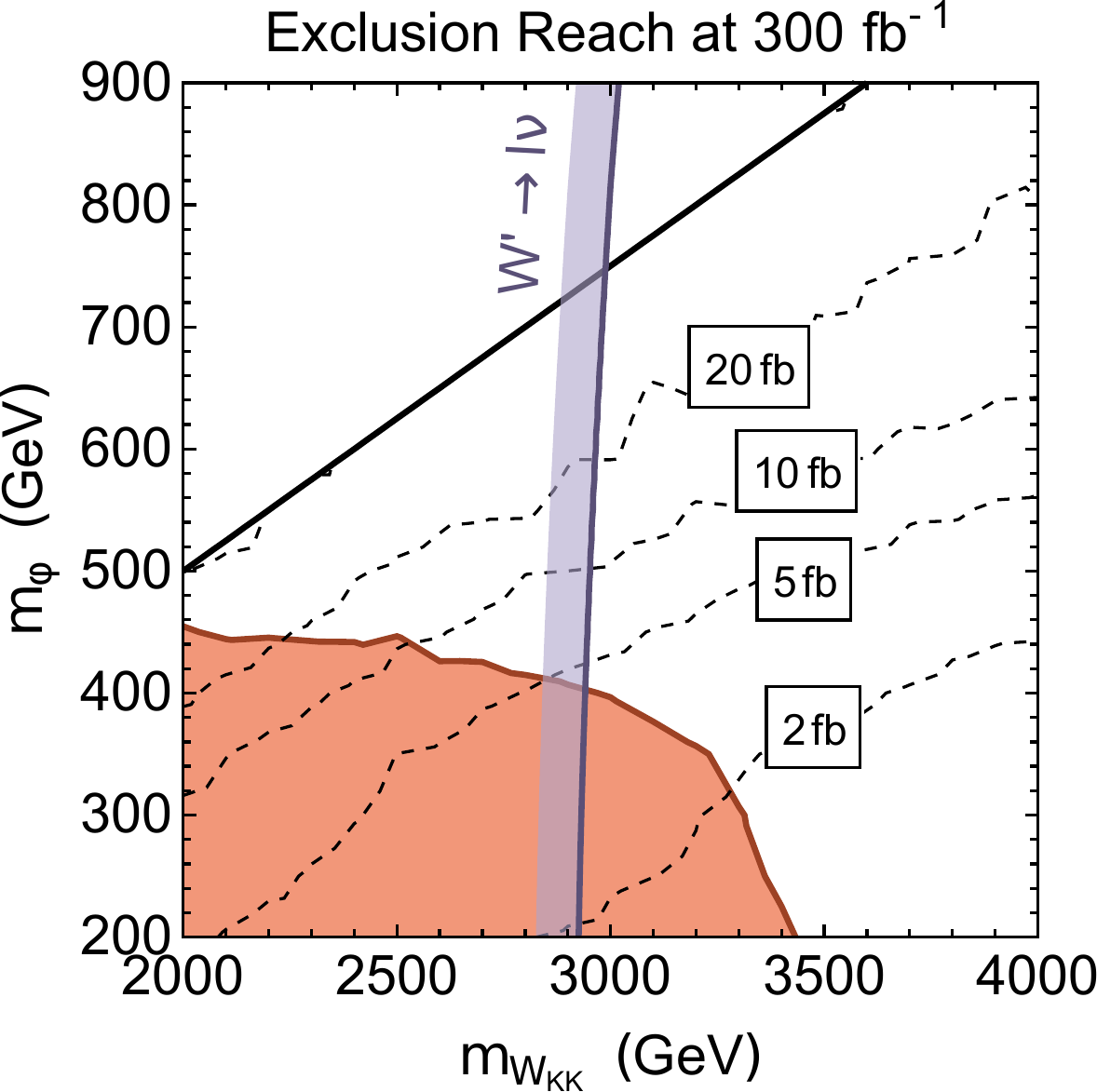}
\caption{\small{Semileptonic diboson jet}}
\label{fig:SemiLeptonic-ExclusionReach}
\end{subfigure}
\begin{subfigure}[b]{0.45\textwidth}
\includegraphics[width=\textwidth]{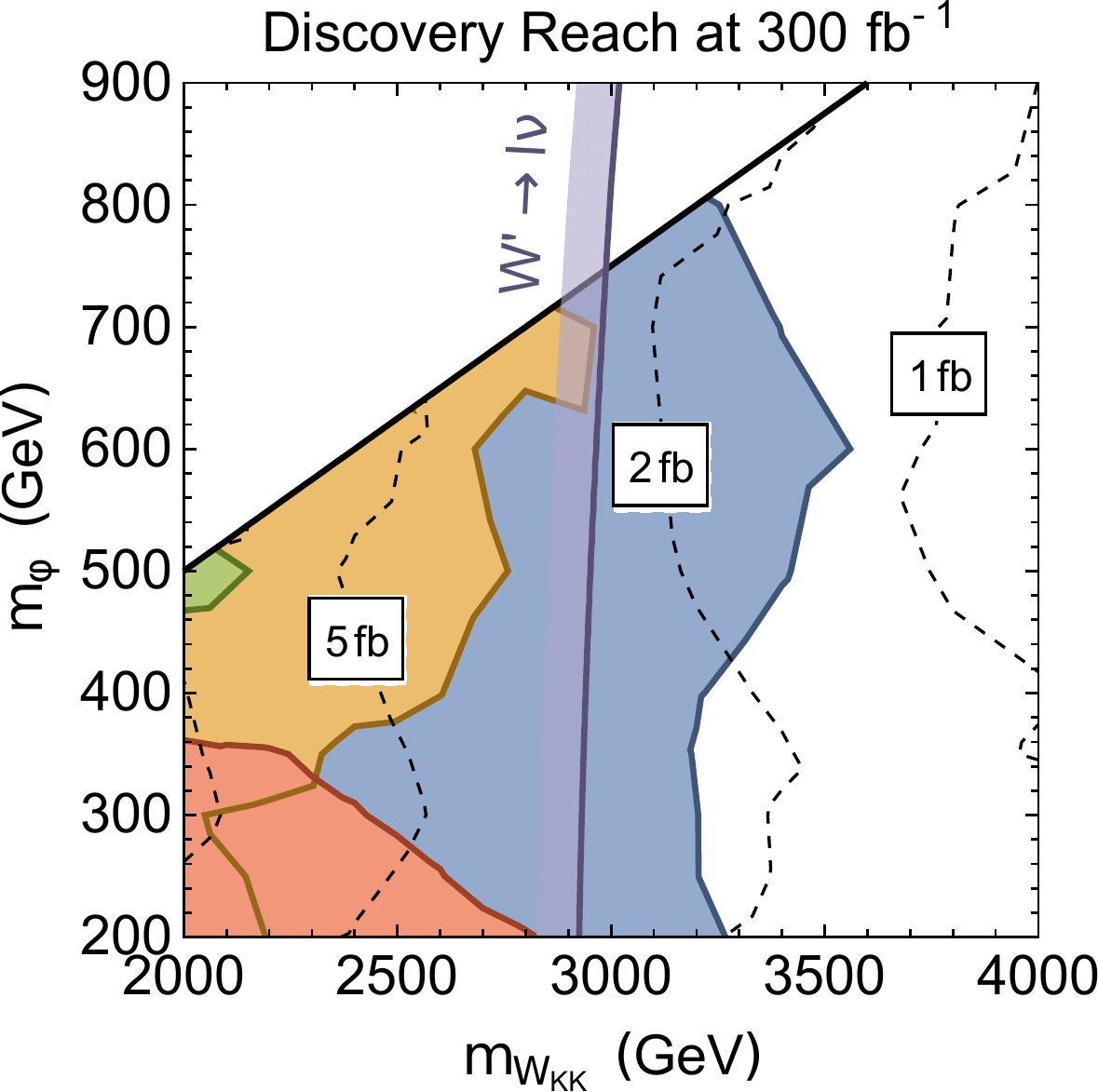}
\caption{\small{Combined discovery reach}}
\label{fig:Combined-DiscoveryReach}
\end{subfigure}
\caption{\small{
Left: Exclusion reach with $\mathcal{L} = 300 \; \text{fb}^{-1}$ for a semi-leptonic diboson jet recoiling against a hadronic $W$ jet. Right: Combined discovery reach with $\mathcal{L} = 300 \; \text{fb}^{-1}$. Red: $qq(\ell \nu qq)$, orange: $\ell \nu (qqqq)$, green: $qq(qqqq)$, blue: combination. Dashed contours indicate the exclusion or combined discovery cross section for $W_\text{KK} \to W \varphi \to W(WW)$. The purple line indicates the current exclusion due to the search for $W' \to \ell \nu$ of Ref.~\cite{ATLAS-CONF-2018-017}.}
}
\label{fig:SemiLeptonic-ExclusionAndDiscoveryReach}
\end{figure}

Using the signal isolation strategy, we get $\mathcal{O}(30)$ signal events and $\mathcal{O}(5)$ background events at $300$ fb$^{-1}$ for low radion and $W_{\text{KK}}$ masses, where the strategy is most effective. We present our exclusion results in the $m_\varphi - m_{W_{\rm KK}}$ plane in Fig.~\ref{fig:SemiLeptonic-ExclusionReach}. The gray dotted lines show the contours for required signal cross section for a $2\sigma$ exclusion while the shaded red region shows the parameter space where this method can be used to probe the extra-dimensional model considered in this work. The purple line indicates the current exclusion due to $W'\rightarrow \ell \nu$ search from Ref.~\cite{ATLAS-CONF-2018-017}. 

\subsubsection{Combination}
\label{sec:warped_results_combination}

The distribution of the $W(WW)$ process into multiple final states with comparable branching fractions and experimental sensitivities presents both a challenge and an opportunity. The challenge is that full coverage requires multiple dedicated searches, each with different SM backgrounds. The opportunity is that a signal in multiple channels simultaneously would provide not only evidence of a new boosted particle, but also evidence of its identity as a resonance decaying into two vector bosons rather than some alternative three- or four-prong topology. We illustrate this in the Fig.~\ref{fig:Combined-DiscoveryReach}. Here we present in green, orange and red the $5\sigma$ discovery potential for the three search channels we have explored using a luminosity of $300\;\text{fb}^{-1}$. With this luminosity, none of these channels individually are likely to provide a discovery for this particular model in light of the existing constraint on the decay $W_\text{KK} \to \ell \nu$ using $80 \; \text{fb}^{-1}$ of data, indicated by the purple line (though this statement is model dependent and might be evaded in a model with a leptophobic parent resonance). However, we also present in blue the discovery sensitivity of a combination of the three searches, along with contours of the discovery cross section for the process  $W_\text{KK} \to W \varphi \to W(WW)$. This has been obtained using the product of likelihood ratios from the individual channels, which in the approximations we are using is equivalent to adding the individual significances in quadrature. While discovery of the $W_\text{KK}$ in this model is likely to be made first in the $\ell \nu$ channel, we see that present constraints allow for the possibility of discovery of both the parent $W_\text{KK}$ and boosted daughter $\varphi$ using jet substructure techniques with $300 \; \text{fb}^{-1}$ of data.

\begin{figure}
\centering
\includegraphics[width=0.6\linewidth]{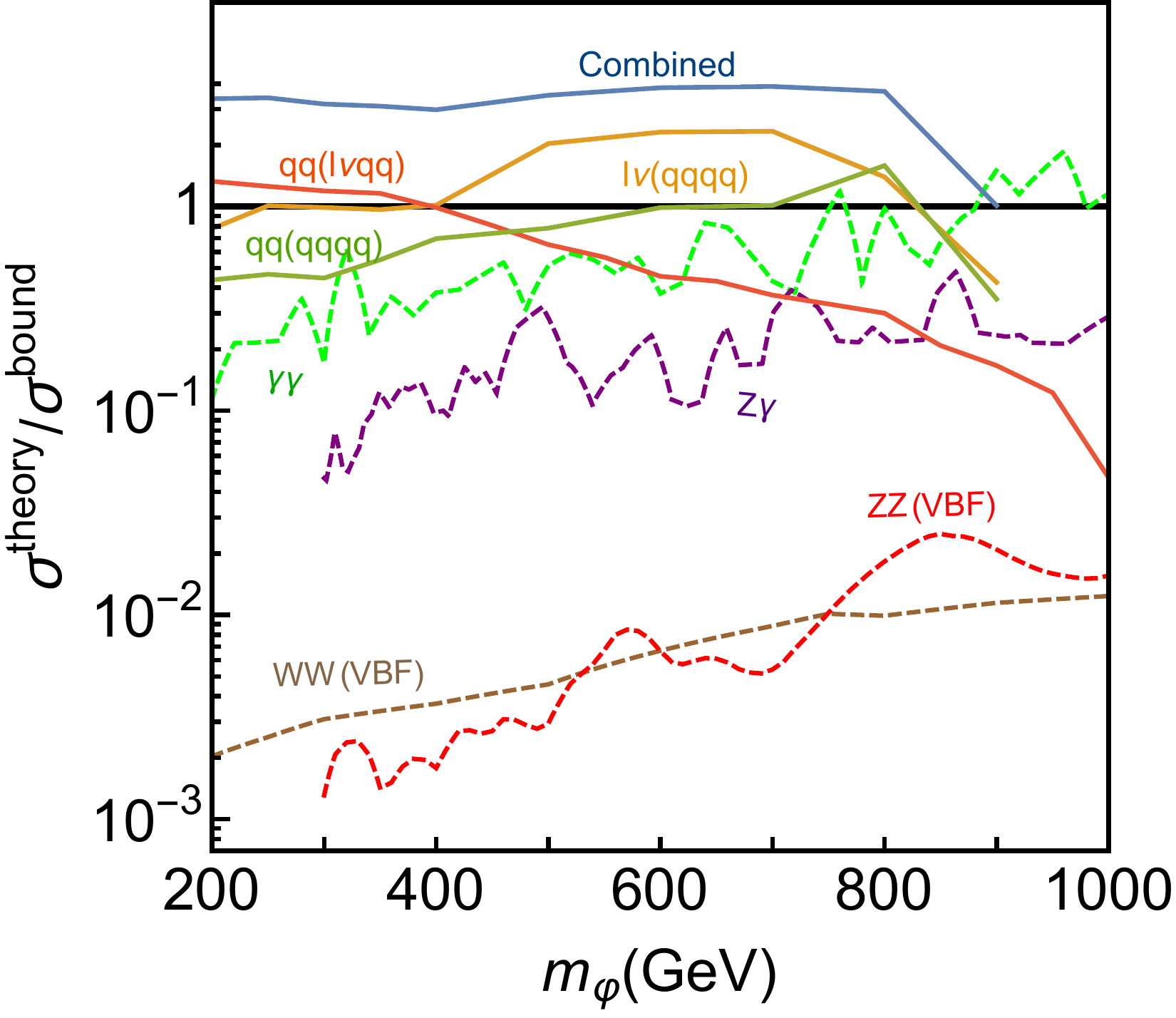}
\caption{
\small{Comparison of search performance for searches dedicated to the direct production of $\varphi$ (dashed) vs searches dedicated for boosted production of a diboson jet (solid), extrapolated to $\mathcal{L} = 300 \; \text{fb}^{-1}$ as described in the text. We have fixed $m_{W_\text{KK}} = 3 \; \text{TeV}$. Cross section limits are reported as a fraction of the predicted cross section in the warped model benchmark described in this section.}
}
\label{fig:bound_comparison}
\end{figure}

A central question is whether a light $\varphi$ of this kind is likely to be discovered first in searches dedicated to its direct production with low boost, or in searches dedicated to its boosted production from the decay of a heavier resonance. We address this question in the context of the warped model benchmark in Fig.~\ref{fig:bound_comparison}. We present in solid red, orange and green lines the sensitivity to $W_\text{KK} \to W \varphi \to W(WW)$ of the boosted searches in the final states $qq(\ell \nu qq)$, $\ell \nu (qqqq)$, and $qq(qqqq)$ respectively, and in blue their combination. The dashed lines represent a naive projection of the sensitivities of the searches described in section~\ref{bound} and at the beginning of this section to a luminosity of $300 \; \text{fb}^{-1}$. This is done by rescaling the limits by a factor $(\mathcal{L}_\text{search} / 300 \; \text{fb}^{-1})^{1/2}$, which is appropriate when the limits are dominated by background statistics as is the case for these searches, and assuming that the analysis used will be the same at the higher luminosity as at the lower. We see that for light radion masses, the limits coming from the boosted diboson jet searches are several times more powerful than the diphoton search (which is the most powerful of the searches for the directly produced radion), and their combination is an order of magnitude more powerful. It should be kept in mind that the comparison with the diphoton limits is quite model dependent; in this model, $\varphi$ has a significant branching fraction into diphotons, which might not be true of alternative models. When compared with the limits coming from the diboson searches ($WW$, $ZZ$), we see that the sensitivity in the boosted configuration is several orders of magnitude greater.

\section{Conclusions and Outlook}
\label{conclude}
 
In this paper, we have studied the signals from the production of a boosted light scalar $\varphi$. Our motivation was guided by the generic presence of such scalars in several BSM scenarios. We focused on the case when the scalar coupling to quarks/gluons is small, as compared to $W/Z/\gamma$. Compared to the only scalar in the SM -- the Higgs, it is a logical possibility that a new scalar may have a different hierarchy of couplings to other SM particles (in this work we have taken the scalar to have a suppressed coupling to the Higgs itself, but one can consider the alternative). Focusing on the direct production bounds (using only couplings to SM particles) on such a scalar, we showed that these constraints are rather weak, given the very small production cross-sections due to suppressed couplings to quarks/gluons; we find that the scalar can be as light as $O(100)$ GeV.

Indeed, in a BSM scenario with such a scalar, there can be other heavier particles which can produce the scalar from their decay. The crucial point is that this light scalar is produced with a high boost, and requires developing observational strategies that would take the boost into account and use it judiciously. The above process in fact results in a striking signature -- a pair of $W/Z/\gamma$ from the decay of the boosted scalar get merged and their decay products (where relevant) also get merged.  Ultimately, such a process results in a multi-pronged ``fat'' object, the ``boosted-diboson''. Without a more customized approach, standard searches are not sensitive to it, which was the motivation for the present study. 

The focus of this work has been the study of dedicated algorithms for tagging a boosted diboson jet. For the fully hadronic decay, we have investigated the application of recursive soft drop techniques to isolate the four prongs and reconstruct a pair of $W$ boson candidates, while for the semileptonic decay we have studied the application of lepton subjet fractions to identify hard but non-isolated leptons within the jet. For the fully hadronic diboson jet, we find significance improvements of 3-10 over QCD jet backgrounds depending on the scalar mass, which is similar to that for a pair of well separated but boosted $W$-jets using standard $W$-tagging algorithms. For the semileptonic diboson jet, we find that an LSF cut allows us to eliminate QCD backgrounds, leaving mainly those coming from boosted tops and from $W$-strahlung off of quark jets.

As a concrete example, we applied the boosted diboson taggers to an extension of the standard warped extra dimensional model, where only EW gauge fields and gravity propagate in an additional part of the bulk. In this case, $\varphi$ can be identified with the radion: its hadro-phobic nature derived from it being spatially sequestered from quarks/gluons in the extra dimension. The radion can be produced from the decay of the EW gauge KK particles. We showed that the application of the dedicated taggers is well motivated --- it allows a reach into parts of parameter space not accessible otherwise. Combining the use of these dedicated taggers in the relevant decay modes of the boosted radion allowed a discovery reach at 300 fb$^{-1}$ in parts of parameter space which are inaccessible by application of one of these taggers individually (in the suitable decay channel). We would like to emphasize here that the boosted production mode has higher potential for discovering the radion, as compared to the usual direct one.     

In conclusion, developing and using dedicated methods to look for decay products of boosted particles at the LHC is well motivated due to the increased sensitivity it can provide for discovery. Further, having different methods suitable for different decay modes of the boosted particles is desirable because together they can provide a better reach, as well as a way to probe the underlying nature of the new physics more directly. 

\section*{Acknowledgements}
We would like to thank Douglas Berry, Seung Lee, Sung Hak Lim, Petar Maksimovic, Ian Moult, Manuel Toharia, Nhan Tran, Marco Trovato and Marc Weinberg for comments and discussions. The work of KA, JHC, and PD was supported in part by NSF Grant No.~PHY-1620074 and the Maryland Center for Fundamental Physics. KA  and JC were also supported by the Fermilab Distinguished Scholars Program. The work of SH was supported by NSF Grant No.~PHY-1719877. The work of DK was supported by the Korean Research Foundation (KRF) through the CERN-Korea Fellowship program, and Department of Energy under Grant No. DE-FG02-13ER41976/DE-SC0009913.

\bibliographystyle{utphys}
\bibliography{bibliography}
 
\end{document}